%% file: robert_kunsch_arxiv_v2.tex
\documentclass[A4]{article}
\pdfoutput=1
\usepackage[usenames,dvipsnames,svgnames,table]{xcolor}
\usepackage[T1]{fontenc}

\usepackage{hyperref}

\usepackage[a4]{crop}
\usepackage{amsmath}
\usepackage{graphicx}
\usepackage{soul}
\usepackage{fancyhdr}
\usepackage{mathtools}

\usepackage{glossaries}
\usepackage[capitalise]{cleveref}
\usepackage{enumitem}
\usepackage{natbib}
\usepackage{bm}

\usepackage[english]{babel}
\usepackage{lmodern}
\usepackage{siunitx}
\usepackage{tikz}
\usetikzlibrary{positioning, arrows, shapes}

\newif\iftikzfigs
\tikzfigstrue 

\hypersetup{
    pdfauthor={Sylvain Robert},
    pdftitle={Localizing the Ensemble Kalman Particle Filter},
    colorlinks=true,
    citecolor=black,
    linkcolor=black,%
    linkbordercolor=white,
    citebordercolor=white
}

\newcommand{\figfactor}{0.8}

\newcommand{\figdist}{.75cm} 
\newcommand{\figdistsmall}{.4cm}  
\newcommand{\figdistlong}{1.4cm}   
\newcommand{\lwidth}{1}                

\newcommand{\twopartdef}[4]
{
  \left\{
  	\begin{array}{ll}
			#1 & \mbox{if } #2 \\
			#3 & #4
		\end{array}
	\right.
}

\newacronym{enkf}  {\textsc{e}\textnormal{n}\textsc{kf}}{Ensemble Kalman Filter}
\newacronym{enkpf} {\textsc{e}\textnormal{n}\textsc{kpf}}{Ensemble Kalman Particle Filter}
\newacronym{letkf} {\textsc{letkf}}{Local Ensemble Transform Kalman Filter}

\newacronym{lenkf}  {\textsc{le}\textnormal{n}\textsc{kf}}{Local Ensemble Kalman Filter}
\newacronym{block}{\textsc{block-le}\textnormal{n}\textsc{kpf}}{block-local \textsc{e}\textnormal{n}\textsc{kpf}}
\newacronym{naive} {\textsc{naive-le}\textnormal{n}\textsc{kpf}} {naive-local \textsc{e}\textnormal{n}\textsc{kpf}}

\newacronym{llensf}{\textsc{lle}\textnormal{ns}\textsc{f}} {local-local \textsc{e}\textnormal{ns}\textsc{f}}

\newacronym{pf}{\textsc{pf}}{Particle Filter}
\newacronym{ess}{\textsc{ess}}{Effective Sample Size}

\newacronym{nwp}{\textsc{nwp}}{numerical weather prediction}
\newacronym{sweq}{\textsc{sweq}}{shallow water equation}

\newacronym{crps}{\textsc{crps}}{continuous ranked probability score}

\newcommand{\xbi}{\ensuremath{x^{b,i}}}
\newcommand{\xai}{\ensuremath{x^{a,i}}}
\newcommand{\kk}{\ensuremath{K(P^b)}}
\newcommand{\kp}{\ensuremath{K(\gamma P^b)}}
\newcommand{\kq}{\ensuremath{K((1-\gamma) Q)}}

\begin{document}

\title{Localizing the Ensemble Kalman Particle Filter}


\author{Sylvain Robert and Hans R.~K\"unsch \\ Seminar f\"ur 
    Statistik, ETH Z\"urich, Switzerland}


\maketitle

\begin{abstract}
    Ensemble methods such as the \gls{enkf} are widely used for data 
    assimilation in large-scale geophysical applications, as for example in 
    \gls{nwp}. There is a growing interest for physical models with higher and 
    higher resolution, which brings new challenges for data assimilation 
    techniques because of the presence of non-linear and non-Gaussian features 
    that are not adequately treated by the \gls{enkf}.
    We propose two new localized algorithms based on the \gls{enkpf}, a hybrid 
    method combining the \gls{enkf} and the \gls{pf} in a way that maintains 
    scalability and sample diversity.
    Localization is a key element of the success of \glspl{enkf} in practice, 
    but it is much more challenging to apply to \glspl{pf}.
    The algorithms that we introduce in the present paper provide a compromise 
    between the \gls{enkf} and the \gls{pf} while avoiding some of the problems 
    of localization for pure \glspl{pf}.
    Numerical experiments with a simplified model of cumulus convection based 
    on a modified \acrlong{sweq} show 
    that the proposed algorithms perform better than the local \gls{enkf}. 
    In particular, the \gls{pf} nature of the method allows to 
    capture non-Gaussian characteristics of the estimated fields such as the 
    location of wet and dry areas.
\end{abstract}


\glsresetall

\section{Introduction}
\vspace*{-1pt}
\noindent
In many large-scale environmental applications, estimating the evolution of a geophysical system, such as the atmosphere, is of utmost interest.
\emph{Data assimilation} solves this problem iteratively by alternating between a forecasting step and an updating step.
In the former, information about the dynamic of the system is incorporated, while in the latter, also called \emph{analysis}, partial and noisy observations are used to correct the current estimate.
The optimal combination of the information from these two steps requires an 
estimate of their associated uncertainty.
In statistics, one represents the uncertainty about the state of a system after 
the forecasting step with a prior distribution, and the uncertainty due to the 
observations errors with a likelihood. 
The analysis consists then in deriving the posterior distribution of the 
current state of the system, 
combining the prior distribution and the new observations, which can be done 
with Bayes' rule.

In geophysical applications, such as \gls{nwp}, the dimension of the system is 
extremely large and the forecasting step computationally heavy, therefore the 
focus is on developing efficient methods with reasonable approximations.
Even in the simplest case of a linear system with linear Gaussian observations, the optimal method, namely the Kalman filter \citep{kalman_new_1960,kalman_new_1961}, is difficult to use because of the size of the matrices involved.

Ensemble, or Monte-Carlo, methods, are  elegant techniques to deal with 
non-linear dynamical systems.
They use finite samples, or \emph{ensembles} of \emph{particles}, to represent 
the uncertainty about the state of the system associated with the prior and 
posterior distributions.
The forecasting step consists then simply in integrating each particle according to the law of the system.
Ensemble methods were introduced in the geosciences by the \gls{enkf} of 
\citet{evensen_sequential_1994,evensen_data_2009} as a Monte-Carlo 
approximation of the Kalman filter and have shown great success in practice.
However, the analysis step of all \gls{enkf} methods implicitly relies on the assumption that the prior uncertainty about the state of the system is Gaussian, which is an acceptable approximation in some cases, but is unlikely to hold with highly non-linear dynamics.

\glspl{pf} 
\citep{gordon_novel_1993,pitt_filtering_1999,doucet_smc_2001} are a 
more general 
class of ensemble methods which differ from the \gls{enkf} in the way the 
analysis step is implemented.
They can handle fully non-linear and non-Gaussian systems  and are therefore 
very attractive.
Unfortunately, it has been shown that the number of particles needed to avoid sample degeneracy and collapse of the filter increases exponentially with the size of the problem, in a sense made precise in \citet{snyder_obstacles_2008}. 
Adapting \glspl{pf} for large-scale environmental applications is an active field of research and there are many propositions of algorithms (see \citet{van_leeuwen_particle_2009} for a review, and \citet{van_leeuwen_nonlinear_2010,papadakis_data_2010,ades_exploration_2013,nakano_hybrid_2014} for more recent developments). 
Here we focus on the \gls{enkpf}, introduced in \citet{frei_enkpf_2013}, which consists in a combination of the \gls{pf} with the \gls{enkf}.
Compared to other similar algorithms, the \gls{enkpf} has the distinct 
advantage of being dependent on a single tuning parameter which defines a 
continuous interpolation between the \gls{enkf} and the \gls{pf}.
Moreover, no approximation of the prior distribution or the transition probability is required.

Recently, there has been a tendency towards using physical models with higher and higher resolution.
For example in \gls{nwp}, regional models are starting to be run with a grid 
length of the order of 1 kilometer, which allows to resolve explicitly highly 
non-linear phenomena such as cumulus convection.
In general, with non-linear dynamical systems, the uncertainty after the forecasting step can become highly non-Gaussian.
Therefore there is a growing need for data assimilation methods which can handle non-linear and non-Gaussian systems while being computationally efficient to be applied to large-scale problems \citep{bauer_nature_2015}.
The \gls{enkf} implicitly assumes Gaussian uncertainty while the \gls{pf} requires an exponentially large number of particles. 
The \gls{enkpf} is a compromise between both, but it still requires too many particles for  practical applications.
The main goal of this article is to contribute towards a full solution to the 
non-linear and non-Gaussian large-scale data assimilation problem.

The methods proposed in this article expand on the \gls{enkpf} by introducing localization.
The idea of localizing the analysis was first proposed by \citet{houtekamer1998data} as a device to reduce dimensions and thus to allow for smaller ensemble sizes.
While localization has been widely used within the \gls{enkf} family of 
algorithms, e.g.\ the \gls{letkf} of \citet{hunt_efficient_2007}, applications 
to the \gls{pf} are much rarer.
The reason for this is that \gls{pf} methods introduce a discrete component in 
the analysis by resampling particles, which  breaks the necessary smoothness of 
the fields to be estimated.

In the literature, the main approaches to this problem have been  either 
to avoid resampling altogether, or to correct the introduced 
discontinuities. 
The moment-matching filter of \citet{lei_moment_2011} avoids resampling and can 
be localized straightforwardly as it depends on the first two moments of 
the distribution only.
Attempts to replace resampling by deterministic transport maps from  prior to  posterior distributions are a promising way to reformulate \glspl{pf} such that they can be easily localized \citep{reich_guided_2013}.
An early example of algorithm which keeps resampling while localizing the 
analysis is the \gls{llensf} of 
\citet{bengtsson_toward_2003}, with which our new algorithm share many 
similarities.
In the recent local \gls{pf} of \citet{poterjoy_localized_2016} resampling is 
applied locally by progressively merging resampled particles with prior 
particles, followed by a higher-order correction using a deterministic 
probability mapping.

In this article, we propose two new localized algorithms based on the 
\gls{enkpf}: the \gls{naive} and the \gls{block}. 
In the \gls{naive}, assimilation is done independently at each 
location, ignoring potential problems associated with discontinuities; 
in the \gls{block}, data are assimilated by blocks, whose influence is limited to a neighborhood, and discontinuities are smoothed out in a transition area by 
conditional resampling.
The first method is easier to implement as it mirrors the behavior of the \gls{letkf}, but the second one deals better with the specific problems associated with localized \glspl{pf}.
The localization of the \gls{enkpf} or any \gls{pf} method is highly 
non-trivial, but it can potentially bring remarkable improvement in terms of 
their applicability to large-scale applications  
\citep{snyder_performance_2015}.

The original \gls{enkpf} has been shown in \citet{frei_enkpf_2013} to perform well on the Lorenz 96 model \citep{lorenz_optimal_1998} and other rather simple setups. 
The extensions that we propose in the present paper should allow the algorithm to work on more complex and larger models.
Here, we test the feasibility of our methods with some numerical experiments on 
an artificial model of cumulus convection based on a modified \gls{sweq} 
\citep{wursch_sweq_2014}  and show that we obtain similar or better results 
than the \gls{enkf}.


In Section 2 we briefly review ensemble data assimilation and the 
\gls{enkf}, the \gls{pf} and the \gls{enkpf}.
Then we discuss localization and explain the two new localized \glspl{enkpf} 
algorithms in  Section 3.
The numerical experiments are described and results discussed in 
Section 4 before a few conclusive remarks in Section 5.

\section{Ensemble data assimilation} \label{sec:da}
\vspace*{-1pt}
\noindent
Consider the problem of estimating the state of a system at time $t$, $x_t$, given a sequence of partial and noisy observations $y_{1:t}=(y_1, \dots, y_t)$.
The underlying process $(x_t)$ is unobserved and represents the evolution of the system, described typically by partial differential equations.
The observations are assumed to be conditionally independent given the states and are characterized by the likelihood $l(x_t|y_t)$.
This problem fits in the framework of general state space models and is generally known as \emph{filtering} in the statistics and engineering community, and as \emph{data assimilation} in the geosciences.

Mathematically, the goal is to compute the conditional distributions of $x_t$ given $y_{1:t}$, called the filtering or \emph{analysis distribution} $\pi^a_t$.
There exists a recursive algorithm which alternates between computing the 
analysis distributions and the conditional distributions of $x_t$ given 
$y_{1:t-1}$, called the \emph{predictive} or \emph{background} distribution 
$\pi^b_t$.
In the forecast step, the background distribution at time $t$ is derived from 
the analysis distribution at time $t-1$, using the dynamical laws of the system.
In the \emph{analysis} or \emph{update} step, the analysis distribution at time $t$ is derived from the background distribution at the same time $t$, using Bayes' theorem: $\pi^a_t(x) \propto \pi^b_t(x) \cdot l(x | y_t)$.
However, there is no analytically tractable solution to this recursion except in the case of a discrete state space or a linear and Gaussian system.
In the latter case, the solution is known as the Kalman filter.

One of the problems that arise when trying to apply this theoretical framework to large-scale systems such as in \gls{nwp} is that the forecast step is not given by an explicit equation but comes from the numerical integration of the state vector according to the dynamical laws of the system.
\emph{Ensemble} or \emph{Monte Carlo} methods address this problem by representing the distributions $\pi^b_t$ and $\pi^a_t$ by finite samples or \emph{ensembles} of \emph{particles}: $\xbi_t \sim \pi^b_t$ and $\xai_t \sim \pi^a_t$ for $i=1,\dots,k$, where $k$ is the size of the ensemble.
The forecast step produces the background ensemble members $\xbi_t$ by 
propagating the analysis ensemble members $x^{a,i}_{t-1}$ according to the 
dynamics of the system.
The analysis step, that is the transformation of the background ensemble $(\xbi_t)$ into the analysis ensemble, is however more challenging for ensemble methods.
There are various solutions to this problem, depending on the assumptions about the distribution $\pi^b$ and the observation process and the heuristic approximations that are used.

Henceforth we drop the time index $t$ and consider the analysis step only.
We also assume that the observations are Gaussian and linear with mean $Hx$ and covariance $R$, where $H$ is the observation operator applied on a state vector $x$ and $R$ is a valid covariance matrix. 
We now review the \gls{enkf} and the \gls{pf} in this context and describe the \gls{enkpf}, before discussing in more detail the problem of localization and introducing new algorithms.

\subsection{The \acrlong{enkf}} 
\vspace*{-1pt}
\noindent
If one assumes that the background distribution $\pi^b$ is Gaussian and that the observations are linear and Gaussian, then the analysis distribution $\pi^a$ is again Gaussian with a new mean and covariance given by simple formulae.
All \gls{enkf} methods are based on this result and apply it by ignoring non-Gaussian features of $\pi^b$.
They use the background ensemble to estimate the mean and covariance of $\pi^b$ and draw the analysis sample to match the mean and covariance of $\pi^a$ under Gaussian assumptions.
Square-root filters such as the \gls{letkf} transform the background ensemble 
so that the first and second moments match exactly those of the estimated 
analysis distribution, whereas the stochastic \gls{enkf} applies a Kalman 
filter update with some stochastically perturbed observations to each ensemble 
member.
More precisely, in the stochastic \gls{enkf} an ensemble member from the analysis distribution is produced as follows:

\begin{align}
 \xai = \xbi + \kk ( y - H\xbi + \epsilon^i ),
\end{align}
where $P^b$ is an estimate of the background covariance matrix and $\epsilon^i 
\sim \mathcal{N}(0, R)$ is a vector of observation perturbations.
$K(P)$ denotes the Kalman gain computed using the covariance matrix $P$ and is equal to $PH'(HPH' + R)^{-1}$.
Conditional on the background ensemble $(\xbi)$, $\xai$ is thus normal with 
mean $\xbi +  \kk ( y - H\xbi)$ and covariance $\kk R \kk'$.
A key idea for the \gls{enkpf} is that, conditional on the background ensemble, the analysis ensemble is a balanced sample of size $k$ from the Gaussian mixture
\begin{align}
\sum_{i=1}^k \frac{1}{k} \mathcal{N}\big( \xbi + \kk ( y - H\xbi), \kk R \kk' 
\big), 
\label{eq:enkf_gm}
\end{align}
where balanced sample means that each component of the mixture is selected 
exactly once.

\subsection{The \acrlong{pf}} \label{sec:pf}
\vspace*{-1pt}
\noindent
The \gls{pf} does not make any assumption about $\pi^b$, but applies Bayes' formula to the empirical distribution provided by the background ensemble.
In its simplest version it represents the background and analysis distributions by weighted samples of \emph{particles}, whose weights are updated at each time step by a factor proportional to the likelihood of the observations.
That is, if $\pi^b$ is represented by the weighted sample $(\xbi,\alpha^{b,i})$, then $\pi^a$ is represented by the weighted sample $(\xai,\alpha^{a,i})$, where
$$\xai=\xbi,\quad \alpha^{a,i} = \frac{\alpha^{b,i} \cdot l(\xbi|y)} {\sum_{j=1}^k \alpha^{b,j} \cdot l(x^{b,j}|y)}.$$ 
In the forecast step, weights remain unchanged.
However, when iterated this leads to sample degeneracy, that is the weights are effectively concentrated on fewer and fewer ensemble members.
To avoid this, a resampling step is introduced, where the particles are resampled with probability proportional to their weight.
This means that the analysis sample contains $x^{b,i}$ $N_i$ times where $E(N_i)=k \alpha^{a,i}$ and $\sum N_i = k$.
In this way particles which fit the data well are replicated and the others eliminated, thus allowing to explore adaptively the filtering distribution by putting more mass in regions of high probability.

There are balanced sampling schemes, where $|N_i - k\alpha^{a,i}|<1$, which reduce the error due to resampling as much as possible (discussion of balanced sampling can be found in \citet{carpenter_improved_1999,crisan_particle_2001} or \citet{kunsch_recursive_2005}).
But resampling also has problems with sample depletion if the likelihood values $l(\xbi|y)$ are very unbalanced or the dynamical system is deterministic.
In that case one has to add some kind of perturbations to the analysis particles, but it is not clear how to choose the covariance of this noise.

Using a vector of resampled indices $I$, such that $P(I(i)=j) \propto \alpha^j$ 
and $ \#\{ I(i)=j, i=1,\dots,k \} = N_j$ for all $j$, we can write the \gls{pf} 
algorithm succinctly as follows:

\begin{enumerate}[label=\arabic*., align=left]
\item
 Compute the weights $\alpha^j \propto l(x^{b,j} | y)$. 
\item Choose the vector of resampled indices $I$, such that $P(I(i)=j) \propto \alpha^j$ and $|N_j - k \alpha^j| < 1$. 
\item For $i=1,\dots,k$, set $\xai = x^{b,I(i)}$.
\end{enumerate}

\subsection{The \acrlong{enkpf}} \label{sec:enkpf} 
\vspace*{-1pt}
\noindent
The \gls{enkpf} \citep{frei_enkpf_2013} is a hybrid algorithm that combines the \gls{enkf} and the \gls{pf} with a single parameter $\gamma \in [0,1]$ controlling the balance between both.
Its core idea is to split the analysis in two stages, following the progressive correction principle of \citet{musso2001improving}.
In a nutshell, the algorithm consists in ``pulling" the ensemble members towards the observations with a partial \gls{enkf} analysis using the dampened likelihood $l(x|y)^\gamma$, and then applying a partial \gls{pf} with the remaining part  of the likelihood, $l(x|y)^{1-\gamma}$.
In this way the algorithm can capture some non-Gaussian features of the distribution (by resampling), while maintaining sample diversity.
For any fixed $\gamma>0$,
it does not converge to the true posterior distribution as the 
number of particles tends to infinity,
unless the background distribution is Gaussian. The justification
of the EnKPF is rather that, for non-Gaussian background distributions,
it reduces the variance of the PF at the expense of a small bias.

We now review the derivation of the algorithm briefly but refer to \citet{frei_enkpf_2013} for more detail.
Assuming linear and Gaussian observations, dampening the likelihood with the 
exponent $\gamma$ is equivalent to inflating the error covariance $R$ by the 
factor $\gamma^{-1}$, and it is easily seen that this is also equivalent to 
using the Kalman gain with the original error covariance $R$ and a dampened 
background covariance $\gamma P^b$.
From the Gaussian mixture representation of the \gls{enkf} analysis described in \cref{eq:enkf_gm}, we can see that the first step of the algorithm produces the partial analysis distribution

\begin{align}
 \pi^{\gamma} = \sum_{i=1}^k \frac{1}{k} \mathcal{N}( \nu^{a,i}, Q), \label{eq:Pig}
\end{align}
where
\begin{align}
 \nu^{a,i} &%
 = \xbi + \kp ( y - H\xbi), \\
 Q &%
 = \frac{1}{\gamma} \kp R \kp'.
\end{align}

For the second step, we have to apply Bayes' formula using $\pi^\gamma$ as the prior and $l(x|y)^{1-\gamma}$ as the likelihood.
This has a closed form solution \citep{alspach1972nonlinear}, namely a Gaussian mixture with new centroids $\mu^{a,i}$, covariance $P^{a,\gamma}$, and unequal weights $\alpha^i$: 

\begin{align}
 \pi^a_{\gls{enkpf}} &%
 = \sum_{i=1}^k \alpha^i \mathcal{N}( \mu^{a,i}, P^{a,\gamma} ), \label{eq:pia}
\end{align}
where
\begin{align}
 \mu^{a,i} &%
 = \nu^{a,i} + \kq (y - H \nu^{a,i} ), \label{eq:muai} \\
 P^{a,\gamma} &%
 = \big( I - \kq H \big) Q, \label{eq:Pai}\\
 \alpha^i & 
 \propto \phi\{ y; H \nu^{a,i}, HQH' + R/(1-\gamma) \},
 \label{eq:alphai}
\end{align}
and $\phi\{y; \mu, \Sigma \}$ denotes the density of a Gaussian with mean 
$\mu$ and covariance matrix $\Sigma$ evaluated at $y$.
One can rewrite the equation for the $\mu^{a,i}$ components directly from the background ensemble as:
\begin{align}
\mu^{a,i} &= x^{b,i} + L^{\gamma}(y - Hx^{b,i}),  \quad \text{where }\label{eq:lgam} \\ 
L^{\gamma} &= \kp + \kq \Big(I - H \kp \Big). \notag
\end{align}

The final analysis sample is obtained as a sample from the Gaussian mixture  
(\ref{eq:pia}), which can be done at a computational cost comparable to the 
\gls{enkf}.
A short description of the algorithm is given as follows:

\begin{enumerate}[label=\arabic*., align=left]
\item
 Compute all the $\mu^{a,j}$ as in \cref{eq:muai}.
\item
 Compute all the weights $\alpha^j$. \label{alg:alphai}
\item Choose the vector of resampled indices $I$, such that $P(I(i)=j) \propto 
\alpha^j$ and $|N_j - k \alpha^j| < 1$. \label{alg:resample}
\item For $i=1,\dots,k$: \label{alg:foreach}
 \begin{enumerate}[label=\alph*), align=left]
 \item
  Generate $\epsilon^{a,i} \sim \mathcal{N}(0, P^{a,\gamma})$.
  \label{alg:epsai}
 \item
  Set $\xai = \mu^{a,I(i)} + \epsilon^{a,i}$.
  \label{alg:xai}
 \end{enumerate}
\end{enumerate}

\begin{figure}[!b]
    \center
 \input{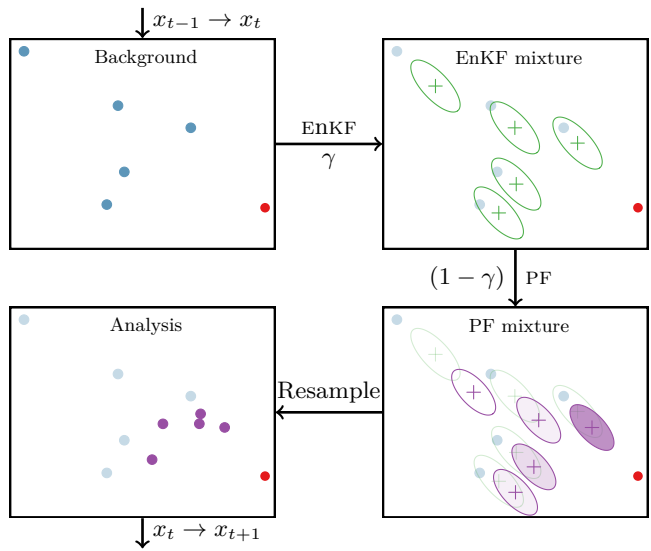}
  \caption[Illustration of the \gls{enkpf}]{Schematic illustration of the
   \gls{enkpf}. Upper left: Background ensemble (blue dots) and
 observation (red dot). Upper right: Intermediate analysis  distribution 
 $\pi^\gamma$ (\ref{eq:Pig}). Each ellipse covers 50\% of one
 component in the mixture. Lower right: Final analysis  distribution
 \cref{eq:pia}. Ellipses again represent 50\% of each component,
 and the color intensity represents the weights $\alpha^i$.
 Lower left: Analysis sample obtained by drawing from \cref{eq:pia}. The 
 mixture
 component closest to the observation has been resam pled 3 times, while the
 two components farthest away have been discarded.} 
  \label{fig:enkpf_illustration}
\end{figure}

The step \ref{alg:foreach}\ref{alg:epsai} can be done efficiently, without computing $P^{a,\gamma}$ explicitly, as described in \citet{frei_enkpf_2013}.
A schematic illustration of the algorithm can be seen in 
\cref{fig:enkpf_illustration}.

In the extreme case of $\gamma=0$, the \gls{enkpf} is equivalent to a pure \gls{pf}, whereas for $\gamma=1$ it is equivalent to the stochastic \gls{enkf}.
$\gamma$ is therefore a tuning parameter which determines the proportion of 
\gls{enkf} and \gls{pf} update to use.
In practice it is chosen adaptively such that the ensemble is as close as possible to the \gls{pf} solution while conserving enough diversity.
Diversity of the mixture weights $\alpha^j$ 
can be quantified by the \gls{ess} \citep{liu_metropolized_1996}.

The \gls{enkpf} has been shown to work well with the Lorenz 96 models and with other simple examples \citep{frei_enkpf_2013}.
However, because it has a \gls{pf} component, it cannot be directly applied to large-scale systems without suffering from sample degeneracy.
In the following section we discuss the technique of localization and introduce two new localized algorithms based on the \gls{enkpf}.

\section{Local algorithms} \label{sec:local}
\vspace*{-1pt}
\noindent 
One of the key element for the success of the \gls{enkf} in practice is localization, either by background covariance tapering as in \cite{hamill_distance_2001}, or by doing the analysis independently at each grid point, using only nearby observations, as in \cite{ott_lenkf_2004}. Localization suppresses spurious correlations at long distances and generally increases the statistical accuracy of the estimates by reducing the size of the problem to solve. Its drawback, however, is that it can easily introduce non-physical features in the global analysis fields. 
For the \gls{lenkf} such problems are reduced by ensuring with some means that the analysis varies smoothly in space.
It should be noted that physical properties of the global fields cannot be guaranteed in a strong sense without incorporating some explicit constraints.

Without Gaussian assumptions localization becomes even more crucial but also more difficult, as the analysis does not anymore depend on the background mean and covariance only. 
The collapse of the \gls{pf} with small ensemble sizes could be avoided by using a very strong localization.
However, a pure local \gls{pf} would probably not be practical as it would introduce arbitrarily large discontinuities in the analysis since different particles can be resampled at neighboring grid points and need to be \emph{glued} together.

Localizing the \gls{enkpf} is easier than for a pure \gls{pf} but it still requires some care due to the resampling step.
We propose two different localized algorithms based on the \gls{enkpf}: the first one is based on the same principle as the \gls{lenkf} of \cite{ott_lenkf_2004}, while the second one is closer to the idea of covariance tapering of \cite{hamill_distance_2001} and serial assimilation of \cite{houtekamer_sequential_2001}, but adapted to the \gls{pf} context.

\subsection{The \acrlong{naive} }
\vspace*{-1pt}
\noindent
In the \gls{naive}, we apply the exact same approach as in the \gls{lenkf} of \cite{ott_lenkf_2004} and do an independent analysis at each grid point.
We call the resulting algorithm \emph{naive} because it ignores dependencies between grid points.
More precisely, the analysis at a given site is produced by sampling from a
local analysis distribution which has the same form as \cref{eq:pia},
but which is computed only using observations close to this site.
As in the \gls{lenkf}, some level of smoothness is ensured by using the same perturbed observations at every grid point and by choosing a local window large enough such that the observations assimilated do not change too abruptly between neighboring grid points.

In order to mitigate
the problem of discontinuities further,
we introduce some basic dependency by
using a balanced sampling scheme with the same random component for every
grid point and by reordering the resampling indices such that the
occurrence of such breaks is minimized. This     
does not remove all discontinuities, but essentially limits them 
 to regions where the resampling weights of the particles change 
quickly.

In conclusion, the \gls{naive} has the advantage to be straightforward to 
implement, following closely the model of the \gls{lenkf}, but it is not 
completely satisfactory as it introduces potential discontinuities in the 
global analysis fields. We now consider a second algorithm which is a bit more 
complicated but avoids this problem.

\subsection{The \acrlong{block}}
\vspace*{-1pt}
\noindent
In the \gls{naive}, localization consists in doing a separate analysis at each grid point, using the observations at nearby locations.
We now consider another approach to localization, in which the influence of each observation is limited to state values at nearby locations.
This seemingly innocuous change of perspective leads to the development of a new algorithm, the \gls{block}.
Assuming that $R$ is diagonal or block-diagonal, the observations $y$ can be partitioned into disjoint blocks $y_1,\dots,y_B$ and then assimilated sequentially, as for example in the \gls{enkf} of \citet{houtekamer_sequential_2001}.
The way that localization is implemented for the \gls{block} is similar in spirit to the global-to-local adjustment of the \gls{llensf} of 
\citet{bengtsson_toward_2003}, but the derivation and the resulting algorithms are not identical. 

In the case of the \gls{enkf}, the influence of one block of observations can be limited to a local area by using a tapered background covariance matrix \citep{hamill_distance_2001}.
However, only in the Gaussian case, setting correlations to zero implies 
independence, but for general $\pi^b$ this is not true.
The \gls{pf} and \gls{enkpf} maintain higher-order dependencies by resampling 
particles globally, but with a local algorithm some dependencies will 
necessarily be broken.
The \gls{block} maintains these dependencies when it is possible, but falls back on a conditional \gls{enkf} and implicitly relies on Gaussian assumptions to bridge discontinuities when they are unavoidable.
We now describe in more detail how to derive the algorithm for one block of observations and then discuss the general method and parallel assimilation.

\subsubsection*{Assimilation of one block of observations}
\vspace*{-1pt}
\noindent 
Let us say we partitioned the observations into $B$ blocks and want to assimilate $y_1$.
If we assume that the observation operator is local, then only a few
elements of the state vector influence the block $y_1$ directly (i.e.\ have
non-zero entry in $H$ for a linear operator). 
We denote their indices by $u$ with corresponding state vector $x_u$.
Hereafter we use subscripts to denote subvectors or submatrices.

Let us assume also that we use a valid tapering matrix $C$, for example 
the one induced by the correlation function given in \citet{gaspari_construction_1999}.
We denote by $x_v$ the subvector of elements that do not influence $y_1$, but are correlated with some elements of $x_u$
(i.e.\ correspond to  non-zero entries in the tapering matrix $C$).
Additionally, we define as $x_w$ the subvector of all remaining elements.

The principle of the algorithm is to  first update  $x_u$ with the
\gls{enkpf} while keeping $x_w$ unchanged. In a second step, $x_v$ 
is updated conditionally on $x_u$ and $x_w$, such that potential
discontinuities are smoothed out. If $x_u$ and $x_w$ are 
not only uncorrelated, but also independent, the background 
distribution can be factored as:
\begin{equation}
\label{pi-b-factor}
\pi^b(x_u, x_v, x_w) = \pi^b(x_u) \pi^b(x_v | x_u, x_w) \pi^b(x_w). 
\end{equation}
By construction, only $x_u$ influences $y_1$ so that one can write 
$l(y_1|x_u,x_v,x_w)=l(y_1|x_u)$. Applying Bayes' rule, the analysis
distribution is
$$
\pi^a( x_u,x_v,x_w | y_1)  \propto \pi^a( x_u | y_1) \pi^b( x_v | x_u, x_w) \pi^b(x_w).
$$
A natural way to sample from this distribution goes as follows: \emph{(i)} 
sample $x_u^{a,i}$ from the analysis distribution $\pi^a (x_u|y_1)$, 
\emph{(ii)} keep $x_w^{a,i}=x_w^{b,i}$ unchanged and \emph{(iii)} sample 
$x_v^{a,i}$ from $\pi^b( x_v | x_u, x_w)$, conditionally on $x_u^{a,i}$ and 
$x_w^{a,i}$.
Steps \emph{(i)} and \emph{(iii)} are clear, but \emph{(ii)} 
requires more discussion.

One could assume normality and sample $x_v^{a,i}$ as a random draw from a normal distribution with the conditional mean and covariance computed from the background sample moments. However this would add unnecessary randomness and it is more judicious to sample $x_v^{a,i}$ as a correction to the background ensemble member $x_v^{b,i}$, as is done in the \gls{enkf}.
Using this sampling scheme, we can show that the analysis of $x_v^{b,i}$ conditioned on $x_u^{a,i}$ and $x_w^{a,i}$ is given by the following simple expression:
\begin{align}
x_v^{a,i} = x_v^{b,i} + P_{vu}^b (P_{uu}^b)^{-1} (x_u^{a,i} - x_u^{b,i}), 
\label{eq:xvcond}
\end{align}
where the matrix inverse is well defined if a tapered estimate of $P^b$ is 
used, and should be understood as a generalized inverse otherwise.

At first sight it is puzzling that $x_w^{a,i}$ does not appear in the formula, but the correlation between $x_v$ and $x_w$ is present in the background sample $x_v^{b,i}$ and thus does not need to be explicitly taken into account in the analysis. 
Note that $x^{a,i}_u$ depends on $x^{b,I(i)}_u$. 

In cases where $I(i)=i$, the
entire particle $x^{a,i}$ is therefore obtained as a correction of the entire 
particle $x^{b,i}$, according to the original \gls{enkpf} algorithm.
In cases where $I(i)\neq i$, $x_v^{a,i}$ will depend on two background particles $x_u^{b,I(i)}$ and $x_u^{b,i}$ and the analysis relies on additional Gaussian assumptions of the background sample. Formula (\ref{eq:xvcond}) then makes sure that the correlation between $x_u^{a,i}$ and $x_v^{a,i}$ is nevertheless correct. In order to stay as close as possible to the EnKPF, we permute the resampling indices $I(i)$ such that the number of cases with $I(i)=i$ is maximal.
More details about the derivation of the algorithm are provided in 
\cref{ap:block}.

Putting everything together, the assimilation of one block of observations in 
the \gls{block} algorithm can be summarized as follows:

\begin{enumerate}[label=\arabic*., align=left]
    \item
    Compute all the $\mu_u^{a,j}$.
    \item
    Compute all the weights $\alpha^j$.
    \item Choose the vector of resampled indices $I$, such that $P(I(i)=j) \propto \alpha^j$ and $|N_j - k \alpha^j| < 1$.
    \item Permute $I$ such that $\#\{j, I(j)= j \}$ is maximal. 
    \label{alg:permute}
    \item For $i=1,\dots,k$:
    \begin{enumerate}[label=\alph*), align=left]
        \item Generate $\epsilon^{a,i} \sim \mathcal{N}(0, P^{a,\gamma}_{uu})$.
        \item Set $x_u^{a,i} = \mu_u^{a,I(i)} + \epsilon^{a,i}$.
        \item Set $x_v^{a,i} = x_v^{b,i} + P_{vu}^b (P_{uu}^b)^{-1} (x_u^{a,i} - x_u^{b,i})$.
        \item Set $x_w^{a,i} = x_w^{b,i}$.
    \end{enumerate}
\end{enumerate}

\begin{figure}
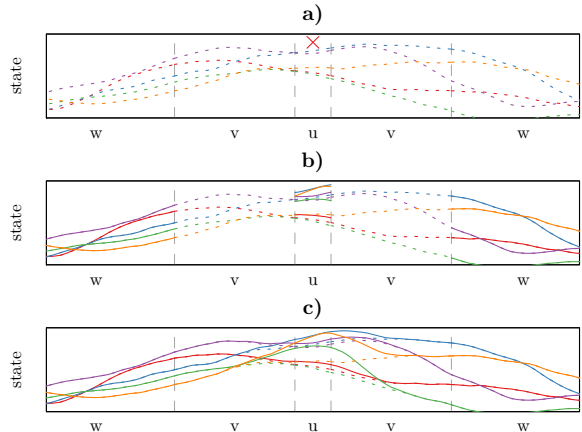

    \centering
    
    \iftikzfigs
    \input{block_illustration.tex}
    \else
    \includegraphics[width=\figfactor \textwidth]{block_illustration}
    \fi
    
    \caption{Illustration of the assimilation of one observation (red cross in panel a)) with the \gls{block}. Each particle is shown in a different color, the dotted lines being the background and the solid lines the analysis. In panel b) $x_u$ is updated while $x_w$ is unchanged. In panel c) we see how the update in $x_v$ makes a transition between $x_u$ and $x_w$. For the orange and green particles, which are not resampled in $x_u$, the analysis has to bridge between two different particles by relying on Gaussian assumptions as described in \cref{ap:block}.}
    \label{fig:block_fig}
\end{figure}

The algorithm is illustrated in \cref{fig:block_fig}.

\subsubsection*{Parallel assimilation of observations}
\vspace*{-1pt}
\noindent
In the previous paragraph we described how one block of observations is assimilated in the \gls{block}.
Now let us consider the case of two blocks of observations to be assimilated, say $y_1$ and $y_2$.
Defining the corresponding state vector indices as above and using an
additional subscript for the block, we can see that if $(u_1,
v_1) \cap (u_2, v_2) = \emptyset$, then in principle
parallel instead of serial
assimilation of $y_1$ and $y_2$ is possible.
In the case where 
$v_1$ and $v_2$ are contiguous, one might worry about discontinuities
at the boundary between $v_1$ and $v_2$. However, the tapering matrix
$C$ ensures that the correlations between sites near this boundary
and sites in $u_1$ and $u_2$ is small and thus the parallel 
assimilation of $y_1$ and $y_2$ through \cref{eq:xvcond} makes
only small changes near this boundary. The procedure could however introduce
some discontinuities in higher-order dependence between $x^{a,i}_{v_1}$
and $x^{a,i}_{v_2}$. To avoid this, one could require an additional buffer
area between $v_1$ and $v_2$, but it would slow down the algorithm and most
likely not bring much improvement. 

We can therefore assimilate all blocks of observations
where the corresponding sets $u$ are well separated in 
parallel. However, blocks where the corresponding sets $u$ are close have
to be assimilated serially. 
In theory, each assimilation of one block increases the 
correlation length because the analysis covariance becomes the new 
background covariance, but we neglect this increase and continue using the
same taper matrix $C$ until all observations have been assimilated.
This additional approximation is necessary to keep the filter local and 
it is also used in the serial \gls{lenkf} of \citet{houtekamer_sequential_2001}.
The resulting algorithm can be described more precisely as follows:

\begin{enumerate}[label=\arabic*., align=left]
    \item
    Partition the observations in $B$ blocks $y_1, \dots, y_B$ and
    determine the sets $u_j$ and $v_j$ ($j=1, \dots, B$).
    \item
    Choose a block $i$ which has not been assimilated so far or exit if none is left.
    \label{alg:choose_b}
    \item
    Assimilate in parallel $y_i$ and all $y_j$ such that $ (u_j, v_j) \cap (u_i, v_i)= \emptyset$. \label{alg:sets}
    \item
    Go to step \ref{alg:choose_b}
\end{enumerate}

The number of times that the algorithm has to loop between successive updates depends on the specific geometry of the problem and on the partitioning of observations in $y_1,\dots,y_B$.
In general one should try to partition the observations such that as many blocks as possible can be assimilated in parallel, but it is not necessary to find the global optimum to this combinatorial problem.

To recapitulate, the \gls{block} consists in assimilating data by blocks and in limiting their influence to a local area. 
The analysis at sites that do not directly influence the observations in the current block but are correlated with $x_u$ is done by drawing from the conditional background distribution.
For cases where resampling does not occur, doing so is equivalent to applying \gls{enkpf} in the local window, whereas for cases where it does occur, the algorithm avoids to introduce harmful discontinuities and produces a smoothed analysis.
The \gls{block} satisfies all our desiderata for a successful localized algorithm based on the \gls{enkpf}.
Its disadvantage, however, is that it requires more overhead for the partitioning of observations and its implementation in an operational setting is more complicated.

Now that we introduced two new localized algorithms based on the \gls{enkpf}, we will proceed to numerical experiments in order to better understand their properties and test their validity by comparing their performance to the \gls{lenkf}.

\section{Numerical experiments} \label{sec:experiments}
\vspace*{-1pt}
\noindent
The algorithms introduced in the present paper can be applied to any task of 
data assimilation for large-scale systems.
However we expect that relaxation of Gaussian assumptions will be most 
beneficial when the dynamical system is strongly non-linear.
Such an application is data assimilation for \gls{nwp} at convective scale.
\citet{wursch_sweq_2014} introduced a simple model of cloud convection which 
allows one to quickly test and develop new algorithms for data assimilation at 
convective scale (as for example in  \citet{haslehner_testing_2016}).
We first briefly introduce the model and the  mechanism to generate artificial 
observations.
Then we present results of two cycled data assimilation experiments.

\subsection{The modified shallow water equation model}
\vspace*{-1pt}
\noindent
The model is based on a modified \gls{sweq} on a one-dimensional domain to 
generate patterns that are similar to the creation of convective precipitations 
in the hot months of summer.
The convective cells are triggered by plumes of ascending hot air generated at random times and locations.
The \gls{sweq} is modified in such a way that if $h$, the height of the fluid, 
in the present case humid air, reaches a given threshold ($h_c$) the convection 
is reinforced and leads to the creation of a cloud.
The convection mechanism is maintained until the fluid reaches a new threshold 
($h_r$), above which the cloud starts to produce rain at a given rate and then 
slowly disappears.
The state of the system can thus be described by three variables: the fluid 
height $h$, the rain content $r$ and the horizontal 
wind speed $u$. We do not use any units as the scales are arbitrary.

The parameters (fluid height thresholds, precipitation rate, etc.) are the same as for the model run described in (\cite{wuersch_testing_2014}, Chap.~5) except 
for the cloud formation threshold set to $H_c=90.02$ as in \citep{wursch_sweq_2014}.
They have been tuned so that the system exhibits characteristics similar to real convection (fraction of clouds, 
life-time of precipitation, etc). The random perturbations are introduced at a rate of $8 \cdot$\num{e-5} \si{\per \meter \per \minute }. We use a domain size of 150 \si{\km} with periodic boundary conditions and a resolution of 500 \si{\meter}.

From this system we generate artificial observations that imitate radar measurements. 
In order to make the experiment realistic, we use a non-linear and non-Gaussian mechanism for generating the observations, but consider them as linear and Gaussian during the assimilation. 
The rain field is observed at every grid point, but set to zero if below a threshold ($r_c$), and with some skewed error whose scatter increases with the average amount of rain otherwise.
Our observation mechanism is different from the one of \citep{wursch_sweq_2014}, where simple truncated Gaussian errors were used, and is intended to be more realistic and challenging than the latter.
In more detail, the rain observations $y_r$ are generated as follows:
\begin{align*}
 y_r = \twopartdef { 0 } { r \leq r_c \text{ or } \frac{1}{2} \epsilon \leq - \sqrt{r-r_c}} { ( \sqrt{r - r_c}+ \frac{1}{2} \epsilon )^2 } {\text{otherwise}},
\end{align*}
where $\epsilon \sim \mathcal{N}(0, \sigma^2_r)$, independently at every grid point.
Such a skewed error distribution for rain observations is a common choice (see 
for example \citet{sigrist_dynamic_2012, stidd_estimating_1973}). It consists 
in applying a Box-Cox transform (with parameter $\lambda=0.5$), adding some 
white noise and then transforming back to the original scale.
Besides rain, wind speed is also observed with some additive Gaussian noise (with variance $\sigma^2_u$), but only at grid points where the observed rain is positive ($y_r \geq r_c$).
For the present experiment $\sigma_r^2=0.1^2$ and $\sigma_u^2=0.0025^2$.

Such artificial observations make data assimilation realistic and challenging due to the non-linearity and sparsity of the observation operator and the non-Gaussian errors.
One could consider transforming the observations to obtain a more normal 
distribution, but we want to test if our algorithms can handle such 
difficult situations.

\begin{figure}
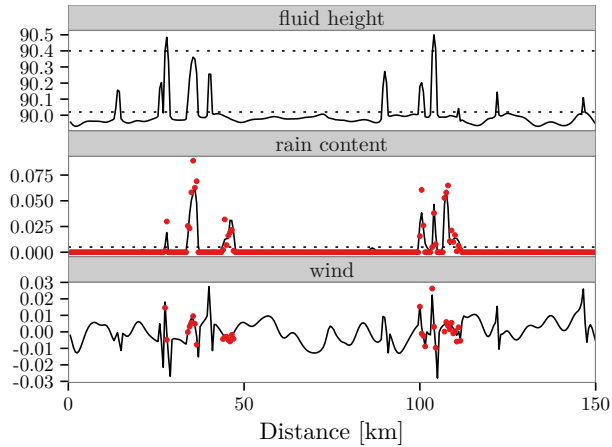

\centering

    \iftikzfigs
     \input{modifiedSWEQ.tex}
    \else
    \includegraphics[width=\figfactor \textwidth]{modifiedSWEQ}
    \fi

 \caption[Illustration of the modified \gls{sweq} model]{Typical example of the modified \gls{sweq} model with artificial radar observations (red 
 dots) and critical values ($h_c$, $h_r$ and $r_c$) as dashed lines.}
 \label{fig:sweq}
\end{figure}

A typical example of a field produced with the model and some artificial 
observations is displayed in \cref{fig:sweq}.
The \emph{bumps} in fluid height represent clouds which start to appear 
if the first threshold is reached (lower dashed line) and are associated with 
precipitation if they reach the second threshold (upper dashed line), after 
which they start to decay.
Rain can remain for some time after a cloud has reached its peak.
The sharp perturbations in the wind field are the random triggering plumes.

\subsection{Assimilation setup}
\vspace*{-1pt}
\noindent 
An initial ensemble of 50 members is generated by letting the model evolve  without assimilating any observations and taking each member at 200 days interval from each other such that they are not correlated. 
We consider only perfect model experiments and do not take into account model error; in particular we do not use any form of covariance inflation.
All the observations are assumed to be Gaussian, with a diagonal covariance matrix $R$  with non-zero elements $R_r$ or $R_u$, depending on which type of observations it corresponds to.
For wind observations the error is the same as for the true generating process, that is we set $R_u=\sigma_u^2$.
Rain and no-rain observations are both assimilated and assumed to have the same  error. 
The true rain distribution is non-Gaussian, so $R_r$ is not straightforward to choose, especially because it depends on the rain level. 
We set $R_r=0.025^2$ during the assimilation, which is equivalent to the error observed for a rain level of 0.06125 (relatively big, but in the range of observed values). 
In general more could be done to treat rain observations properly, but it is beyond the scope of the present study.

We use one localization parameter $l$ set to 5 \si{\km}. 
Every method uses a taper of the covariance matrix as defined in \citet{gaspari_construction_1999} with half correlation length $l$.
For \gls{lenkf} and \gls{naive}, the size of the local window is set to $l$ 
in each direction, for a total of approximately 10 \si{\km} or 21 grid points.
Similarly, the observation blocks for the \gls{block} are defined from segments 
of 10 \si{\km} in the domain (one block contains all the rain observations 
falling in a specific 10 \si{\km} segment and the associated wind observations, 
if any).

The \gls{enkpf} has one free parameter, $\gamma$, which controls the balance between  \gls{enkf} and \gls{pf}. We choose it adaptively such that the \gls{ess} at the resampling step is between 50 and 80\% of the ensemble size. A different $\gamma$ can be selected for each site in the case of the \gls{naive}, or for each block of observations for the \gls{block}, which allows the method to  be closer to the \gls{pf} in regions where non-Gaussian features are present and to fall back closer to the \gls{enkf} when it is necessary. In general the criterion for adaptive $\gamma$ could be refined and tuned more closely, but it is beyond the scope of the present paper.

The two new local algorithms are compared against the \gls{lenkf} of 
\citet{ott_lenkf_2004} and not the \gls{letkf} of \citet{hunt_efficient_2007}, 
because both the \gls{enkpf} and the \gls{lenkf} are based on the stochastic 
\gls{enkf} and thus are more comparable. Furthermore our results cannot be 
directly compared to the ones in  \citet{haslehner_testing_2016} as our 
experimental setup is substantially different from theirs.

\subsection{Results}
\vspace*{-1pt}
\noindent 
In order to highlight some key properties of the new proposed algorithms, we start with an example where high-frequency observations are assimilated and study the resulting analysis ensembles visually.
In a second step, repeating this experiment many times, we can evaluate the 
performance of the algorithms and their differences. 
In a third step we discuss longer assimilation periods with lower frequency 
observations.
We show the results as figures only as we believe that they are only indicative of some possible advantages but should not be taken too literally as the system under study is very artificial and the results can vary with different choice of parameters. 
The quality of assimilation is assessed with the \gls{crps} \citep{gneiting_strictly_2007} commonly defined as:
$$
\text{CRPS}(F, x) = \int ( F(x') - 1_{(x' \geq x)} dx'),
$$
where $F(\cdot)$ is the predictive cumulative probability function, in our case given by the empirical distribution of the ensemble. Because we have a perfect model scenario we can directly evaluate the \gls{crps} of the one-step ahead forecast ensemble compared to the underlying true state of the system. 
The \gls{crps} is a strictly proper scoring rule, which implies that using such a score allows one to control calibration and sharpness at the same time, contrary to the more commonly used \textsc{rmse} (see \citet{gneiting_probabilistic_2014} for a general discussion of probabilistic forecasting).


\subsubsection*{High-frequency observations}
\vspace*{-1pt}
\noindent 
In this first scenario we are interested in seeing if it is possible to use high-frequency radar data, especially for short term prediction. To do so we run a cycled experiment where data are assimilated every 5 \si{\minute} for a total of 1 hour.
Starting from an initial ensemble which has no information about the current meteorological situation, the goal of the filter is to quickly capture areas of rain from the observations. 

\begin{figure}
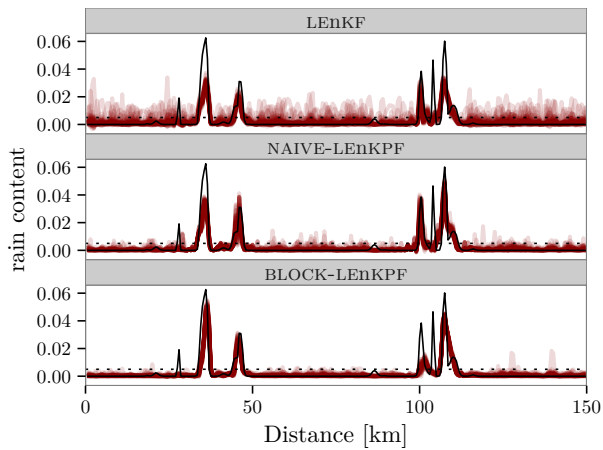

    \centering
    \iftikzfigs
    \input{sweq_rain_filter.tex}
    \else
    \includegraphics[width=\figfactor \textwidth]{sweq_rain_filter}
    \fi
    
    \caption[Illustration of analysis ensembles for the rain field.]{
        Typical example of analysis ensembles for the rain field after 1 hour of high-frequency observations assimilation. Each of the red line is one ensemble member. On top is the \gls{lenkf}, followed by the \gls{naive} and the \gls{block}.}
    \label{fig:rain_filter}
\end{figure}

The analysis ensembles of the different algorithms for the rain field show that the local \glspl{enkpf} are better able to identify dry areas. 
In \cref{fig:rain_filter} we can see the analysis ensembles after one hour of 
assimilation in the same typical situation as in \cref{fig:sweq}.
All methods recover the zones of heavy precipitation relatively well, with some minor differences in terms of maximum intensity. 
One should not conclude too much from an isolated case, but there
is one interesting trait which is not peculiar to this example and illustrates how the local \gls{enkpf} algorithms are able to model non-Gaussian features: the \gls{lenkf} maintains some medium level of rain at almost all sites, while both \gls{naive} and \gls{block} are better at using the \emph{no-rain} observations to suppress spurious precipitation.

Going beyond this particular example, we now consider a simulation study where 
we repeated the above experiment 1000 times and computed the average \gls{crps} 
of each algorithm for every assimilation cycle. To make the results more 
understandable and to remove some natural variability, we always compute the 
performance relative to a free forecast run. The latter is based on the same 
initial ensemble used by all algorithms, but does not assimilate any observation, and is thus equivalent to a climatological forecast.  

\begin{figure}
    \centering 
    \iftikzfigs
    \input{HF_crps.tex}
    \else
     \includegraphics[width=\figfactor \textwidth]{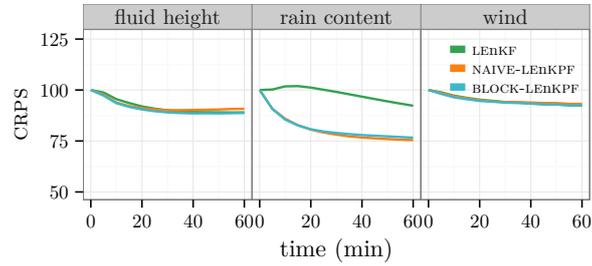}
    \fi
         
    \caption[Evolution of the \gls{crps} for high-frequency data assimilation.]{
        Evolution of the \gls{crps} in the first hour with high-frequency observations. The value is given as a percentage relative to a free forecast run. Notice the truncated y-axis.
        }
    \label{fig:crps_hf}
\end{figure}

Both new algorithms achieve good performances compared to the \gls{lenkf} for the first hour of assimilation, as can be seen in \cref{fig:crps_hf}. 
The gains are in terms of rain content, probably because it is the field with the most non-Gaussian features. 
The \gls{block} seems to have a slight advantage over the \gls{naive} for the fluid height and the other way round for the rain field, but otherwise their performance is very similar.

\begin{figure}
    \centering 
    \iftikzfigs
    \input{HF_U-pattern.tex}
    \else
    \includegraphics[width=\figfactor \textwidth]{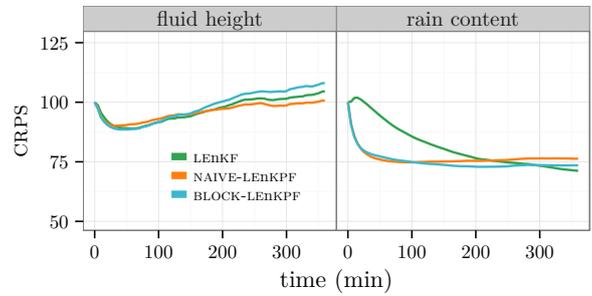}
    \fi
    \caption[U pattern in \gls{crps} evolution for high-frequency observations.]{
        Evolution of the \gls{crps} (relative to a free forecast run) for the  
        fluid height and rain fields in the first 6 hours with high-frequency 
        observations. It becomes obvious that after the first initial 
        improvement, all algorithms deteriorate in term of their ability to 
        capture the underlying fluid height. Notice the truncated y-axis.  
    }
    \label{fig:u_pattern}
\end{figure}

Interestingly, if one looks at the evolution of the \gls{crps} for an 
assimilation period of six hours instead of only one hour in \cref{fig:u_pattern}, some issues start to 
become apparent. 
After an initial drop, the \gls{crps} for the fluid height field increases again and gets worse than the free forecast reference, which means that observations actually hamper the algorithms. 
The effect is greatest for the fluid height but also slightly visible for the rain for the \gls{naive}. 
Physically, the problem comes from the fact that there is a delay between the formation of a cloud and the appearance of rain. Having assimilated many  no-rain observations the algorithm can become overconfident that an area is dry and cannot adapt when new rain starts to appear. 
The \gls{lenkf} is a bit less susceptible to this problem, for the simple reason that it is less good at identifying dry areas and thus maintains more spread in the ensemble while the local \glspl{enkpf} are too sure that no rain is present. Such an effect can be understood as a form of sample degeneracy coming from the fact that a large number of observations are assimilated, which will be confirmed in the next experimental setup. 

\begin{figure}
    \centering
    \iftikzfigs
    \input{calibration.tex}
    \else
    \includegraphics[width=\figfactor \textwidth]{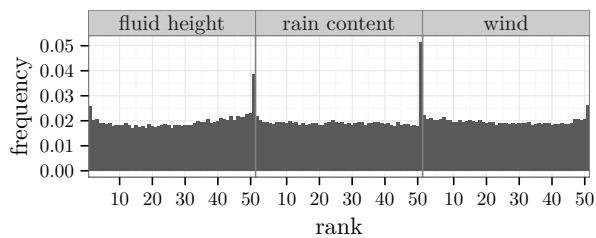}
    \fi
    \caption[Rank histograms.]{
        Rank histograms computed at one-step ahead forecast for the \gls{block} 
        in the high-frequency observations experiment.
        Only every 10 grid points and every 30 minutes are used to increase 
        independence between observations.}
    \label{fig:calibration}
\end{figure}

To assess the calibration of the algorithms we also look at the rank histograms of all fields in the first hour of assimilation for the \gls{block} in \cref{fig:calibration}.
The one-step ahead forecast is more or less calibrated, except for a 
non-negligible fraction of cases where the truth lies outside the range of the 
ensemble.
These problems can be attributed to the inherent difficulty coming from the fundamentally random nature of the system. 
Indeed, attempts at improving the calibration with covariance inflation and 
tuning of the $R$ matrix have not been successful.
The histograms for the fluid level and the rain content reveal that some newly 
appeared clouds are missed, while some spurious clouds are sometimes created. 
The histogram for the absolute wind speed is generally uniform except for a 
slight underestimation, which comes from the  random perturbations of the wind 
field. Conclusions are similar for other algorithms and no clear 
differences can be identified. Therefore the improvements in \gls{crps} can be 
interpreted mainly as better sharpness while keeping calibration the same. 

\subsubsection*{Low-frequency observations}
\vspace*{-1pt}
\noindent 
In a second scenario we consider the assimilation of lower frequency 
observations (every 30 \si{\minute}) but for a longer period  (three days). It would be in principle possible to run one long cycled experiment and to compute the average performance of the algorithms, but we decided to run 100 repetitions of a three days assimilation period instead, because it can be done in parallel and it is more fault tolerant. 

\begin{figure}
    \centering
     \iftikzfigs
         \input{LF_boxplot.tex}
     \else
         \includegraphics[width=\figfactor \textwidth]{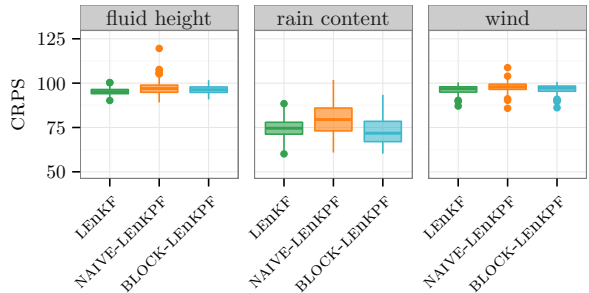}
     \fi

    \caption[Boxplot of the \gls{crps}]{
        Low-frequency observations assimilated for a period of 3 days. Boxplot of the \gls{crps} for the different algorithms and fields considered. The values are given relative to a free forecast. Notice the truncated y-axis.
        } \label{fig:lf_boxplot}
\end{figure}

In this scenario the U-shape pattern highlighted in \cref{fig:u_pattern} is not present anymore, as more diversity is introduced between each assimilation cycle and the ensemble does not become overconfident. 
In term of calibration, the results are similar to the high-frequency scenario, but with less tendency to create spurious clouds and rain but a slight bias towards too small clouds. 
The boxplots of \cref{fig:lf_boxplot} show that \gls{block} outperforms the other methods for the rain field, while the \gls{naive} shows most difficulties, especially for the fluid height where it gets sometimes worse than the free forecast.
One can notice that the fluid height and the wind fields are the most difficult to capture, which comes as no surprise as they are not observed. Furthermore, it should be noted that there is a lot of variability from experiment to experiment, and the \gls{crps} gets regularly worse than the free forecast. 

%

The relatively less good performance of the \gls{naive} compared to the \gls{block} in this scenario might come from the added discontinuities in the analysis.
In the high-frequency scenario the problem was not apparent as the system only 
evolved for a short time before new observations were assimilated.
With low-frequency observations, however, the discontinuities introduced by the 
\gls{naive} have more time to produce a detrimental effect on the dynamical 
evolution of the system, which results in a poorer performance. 

\section{Summary and discussion} \label{sec:conclusion}
\vspace*{-1pt}
\noindent 
We introduced two new localized algorithms based on the \gls{enkpf} in order to 
address the problem of non-linear and non-Gaussian data assimilation, which is 
becoming increasingly relevant in large-scale applications with higher 
resolution. 
The algorithms that we propose combine the \gls{enkf} with the \gls{pf} in a way that avoids sample degeneracy. 
We took particular care to localize the analysis without introducing harmful discontinuities, which is an inherent problem of local \glspl{pf}.

The results of the numerical experiments with a modified \gls{sweq} model 
confirm that the proposed algorithms are promising candidates  for 
application to convective scale data assimilation problems and have some 
distinct advantages compared to the \gls{lenkf}.
The two local \glspl{enkpf} provide better estimates of the rain field, which has non-Gaussian characteristics and thus benefits greatly from the \gls{pf} component of the algorithms.
This advantage is the strongest either in the high-frequency scenario for
short assimilation periods, or in the low-frequency scenario on longer time scales for the \gls{block}. 
Calibration is not perfect as can be seen from the rank histograms, but the improvements in \gls{crps} indicate that the new algorithms are better in terms of sharpness while keeping calibration the same. 
In general the \gls{naive} performs a little worse than the \gls{block}, but it is more straightforward to implement and might thus be suitable for large scale applications.

Assimilating high-frequency observations over long periods of time seems to be problematic for all algorithms, as they grow overconfident and are not able to adapt when new clouds appear in the field. 
Indeed, if one does not assimilate other types of observations the filter is not able to correctly capture the unobserved fluid height field before it produces rain and performances deteriorate quickly after an initial improvement.
It is not certain if such a behavior is particular to the present \gls{sweq} 
model or if it is an inherent characteristic of convective scale assimilation, 
but it indicates some potential limits of such scenarios.
One possible path to tackle this issue would be to properly account for model error. Indeed, the present case study is not a perfect model experiment, as we assume that the observations are linear and Gaussians where in fact they are not. The typical approach to account for model error is to inflate the covariance, but there is a lot of research to do to understand how to apply such ideas in the context of \gls{pf} like algorithms. 

In many applications, the use of a square-root filter such as the \gls{letkf} 
has been shown to be of great benefit. 
Therefore, we are currently investigating possibilities to reformulate the \gls{enkpf} in this framework.
In order to study the impact of localization on the quality of the analysis, we applied the two new localized \glspl{enkpf} to some simpler setups than in the present paper, where it is possible to analyze more closely the problem of discontinuities \citep{robert_arxiv_baysm_2016}.
Given the promising results of the localized \glspl{enkpf}, we are  collaborating with Meteoswiss and Deutscher Wetterdienst on adapting our algorithms to 
\textsc{cosmo}, a convective scale, non-hydrostatic \gls{nwp} model. 


\section{Acknowledgments}
We are very thankful to Daniel Leuenberger from Meteoswiss for fruitful discussions and comments, to Michael W\"{u}rsch for his support for implementing and interpreting his modified \gls{sweq} model, and to the reviewers for their valuable comments.

\appendix 
\section{details for BLOCK-LEnKPF}
\label{ap:block}
Using the factorization (\ref{pi-b-factor}) and $l(y_1|x_u, x_v,
x_w)=l(y_1|x_u)$, Bayes' rule gives the following factorization of the
analysis distribution
\begin{align}
\pi^a(x_u,x_v,x_w | y_1) = \frac{l(y_1|x_u) \pi^b(x_u) \pi^b(x_w) 
\pi^b(x_v|x_u,x_w)
  }{ \pi^b(y_1)}
\end{align}
where
\begin{align*}
 \pi^b(y_1)= \int l(y_1|x) \pi^b(x)  dx =
\int  l(y_1|x_u) \pi^b(x_u) dx_u.
\end{align*}
By integrating out $x_v$, we obtain
\begin{align}
\pi^a(x_u, x_w | y_1) &= \int \pi^a(x_u, x_v, x_w | y_1) dx_v \notag \\
	 &= \frac{l(y_1|x_u) \pi^b(x_u) }{\pi^b(y_1)} \pi^b(x_w) \cdot 1 \notag\\
         &= \pi^a(x_u | y_1) \pi^b(x_w). 
\end{align}
The last step can be easily checked by integrating out $x_w$ and $x_u$
respectively.  
The posterior of $x_w$ is nothing else than the prior, which comes as no
surprise because $x_w$ is independent from $x_u$ and thus is not influenced
by $y_1$. Also, independence between $x_u$ and $x_w$ continues to hold.

Finally, the conditional posterior of $x_v$ can be derived from
the definition of conditional probability 
\begin{align}
\pi^a(x_v | x_u, x_w, y_1) &= \frac{\pi^a(x_u, x_v, x_w | y)}{\pi^a(x_u, x_w | 
y_1)} \notag \\
&= \frac{l(y_1|x_u) \pi^b(x_u) \pi^b(x_w) \pi^b(x_v |x_u,x_w)}
{l(y_1|x_u) \pi^b(x_u) \pi^b(x_w)} \notag \\
					    &= \pi^b(x_v | x_u, x_w), 
\end{align}
where we see that all the terms involving $y_1$ cancel out. Therefore the
conditional posterior of $x_v$ is nothing else than the conditional prior
distribution. 

Assuming additionally that $\pi^b$ is normal, the conditional distribution 
of $x_v$ is again normal with the following mean and covariance:
\begin{align}
\mu^{b}_{v |  u w} &%
= \mu_v^b +  M_{vu} (x_u - \mu_u^b) + M_{vw} (x_w - \mu_w),  \\
P^{b}_{v |  u w} &%
= P^b_{vv} - M_{vu} P^b_{uv} - M_{vw} P^b_{wv}, \\
\text{where } \notag \\
M_{vu} &=  P^b_{vu} (P_{uu}^{b})^{-1} \; \text{and} \;  M_{vw} = 
P^b_{vw}  (P_{ww}^b)^{-1}.
\end{align}
However, instead of making a new random draw from this conditional normal
distribution, one can instead devise a method which reuses the background
samples. Again under the Gaussian assumption, the residual
\begin{equation} 
r_v^{b,i} = x_v^{b,i} - \big(\mu_v^b + M_{vu} (x_u^{b,i} - \mu_u^b) + 
M_{vw} (x_w^{b,i} - \mu_w^b) \big)
\end{equation}
is independent of $x^{b,i}_u$ and $x^{b,i}_w$ and normally distributed
with mean $0$ and covariance $P^{b}_{v |  u w}$. Hence we can use this
residual for sampling from the conditional distribution  
$\pi^a(x_v| x_u^{a,i}, x_w^{a,i}, y_1)=\pi^b(x_v| x_u^{a,i}, x_w^{a,i})$:
\begin{align}
x_v^{a,i} &= \mu_v^b + M_{vu} (x_u^{a,i} - \mu_u^b) + 
M_{vw} (x_w^{a,i} - \mu_w^b) + r_v^{b,i}\notag \\
&= x^{b,i}_v + M_{vu}(x_u^{a,i} - x_u^{b,i})
\end{align}
because $x_w^{a,i}=x_w^{b,i}$.  
Plugging in the definition of $M_{vu}$ we obtain the analysis of 
\cref{eq:xvcond}.

In order to understand better how the \gls{block} works, let us compare it
to the \gls{enkpf} analysis of the entire state $x=(x_u,x_v,x_w)$ using the 
block $y_1$ and not assuming the factorization \cref{pi-b-factor}. Applying the definitions \cref{eq:lgam} and \cref{eq:Pai}
in the case where $P^b$ and $H$
have the block structure
\begin{equation}
P^b = \left(\begin{array}{ccc} P^b_{uu} & P^b_{uv} & 0\\
P^b_{vu} & P^b_{vv} & P^b_{vw}\\ 0 & P^b_{wv} & P^b_{ww} \end{array}
\right)
\end{equation}
and 
\begin{equation}
H=(H_u \ \ 0 \ \ 0) 
\end{equation}
it can be easily verified that
\begin{equation}
L^\gamma = ({L^\gamma_u}' \ \ (M_{vu} {L^\gamma_u})' \ \ 0)'
\end{equation}
and 
\begin{equation} 
P^{a,\gamma} = \left(\begin{array}{ccc} P^{a,\gamma}_{uu} & 
P^{a,\gamma}_{uu} M_{vu}' & 0\\
M_{vu} P^{a,\gamma}_{uu} & M_{vu} P^{a\gamma}_{uu} M_{vu}' & 0\\ 
0 & 0 & 0 \end{array} \right).
\end{equation}
Therefore
\begin{align} 
\mu^{a,i}_v &= x^{b,i}_v + M_{vu}L^\gamma_u(y-H_u x^{b,i}_u) \\
            &= x^{b,i}_v + M_{vu}(\mu^{a,i}_u -x^{b,i}_u)
\end{align}
and
\begin{equation}
\epsilon^i_v = M_{vu} \epsilon^i_u.
\end{equation}
Moreover $\mu^{a,i}_w = x^{b,i}_w$ and $\epsilon^i_w=0$. Combining these
results with  \cref{eq:pia}, we obtain the following for the analysis of $x^{a,i}_v$ with a full \gls{enkpf}:
\begin{equation} 
x^{a,i}_v = x^{b,I(i)}_v + M_{vu}(x^{a,i}_u - x^{b,I(i)}_u),
\quad x^{a,i}_w = x^{b,I(i)}_w. \label{eq:xvai-enkpf}
\end{equation}
Therefore the \gls{block} update given by \cref{eq:xvcond} is the same as the 
\gls{enkpf} update given by \cref{eq:xvai-enkpf} for those indices where $I(i)=i$. However, for indices $i$ 
with $I(i) \neq i$, the \gls{enkpf} analysis applies a correction to 
$x_v^{b,I(i)}$ and not to
 $x_v^{b,i}$ and the size of the correction depends on 
$x^{a,i}_u- x^{b,I(i)}_u$ and not on $x^{a,i}_u-x^{b,i}_u$.
Moreover $x^{b,i}_w$ is replaced by $x^{b,I(i)}_w$ which is in
 conflict with the requirement of a local analysis. Applying a correction
to $x^{b,I(i)}_v$ while setting $x_w^{a,i}=x_w^{b,i}$ would introduce a
discontinuity between $x^{a,i}_v$ and $x^{a,i}_w$. Therefore if 
$I(i)\neq i$, we do not apply an \gls{enkpf} analysis to $x_v^{b,i}$, but
use instead 
 \cref{eq:xvcond}, ensuring a
smooth transition between $x^{a,i}_u$ and $x^{b,i}_w$. If $x^{b,i}_u$ and
$x^{b,i}_w$ are only uncorrelated, but not independent, we ignore
some higher order dependence between these values by pairing
$x^{a,i}_u$ with $x^{b,i}_w$ in cases where $I(i) \neq i$, but this
seems unavoidable.

\bibliography{tellus_bib}{} \bibliographystyle{chicago}

\end{document}

%% file: HF_crps.tex
\begin{tikzpicture}[x=1pt,y=1pt]
\definecolor{fillColor}{RGB}{255,255,255}
\path[use as bounding box,fill=fillColor,fill opacity=0.00] (0,0) rectangle (231.26, 97.56);
\begin{scope}
\path[clip] ( 29.60, 23.76) rectangle ( 92.80, 88.05);
\definecolor{fillColor}{RGB}{255,255,255}

\path[fill=fillColor] ( 29.60, 23.76) rectangle ( 92.80, 88.05);
\definecolor{drawColor}{gray}{0.98}

\path[draw=drawColor,line width= 0.6pt,line join=round] ( 29.60, 36.29) --
	( 92.80, 36.29);

\path[draw=drawColor,line width= 0.6pt,line join=round] ( 29.60, 55.52) --
	( 92.80, 55.52);

\path[draw=drawColor,line width= 0.6pt,line join=round] ( 29.60, 74.75) --
	( 92.80, 74.75);

\path[draw=drawColor,line width= 0.6pt,line join=round] ( 42.05, 23.76) --
	( 42.05, 88.05);

\path[draw=drawColor,line width= 0.6pt,line join=round] ( 61.20, 23.76) --
	( 61.20, 88.05);

\path[draw=drawColor,line width= 0.6pt,line join=round] ( 80.35, 23.76) --
	( 80.35, 88.05);
\definecolor{drawColor}{gray}{0.90}

\path[draw=drawColor,line width= 0.2pt,line join=round] ( 29.60, 26.68) --
	( 92.80, 26.68);

\path[draw=drawColor,line width= 0.2pt,line join=round] ( 29.60, 45.91) --
	( 92.80, 45.91);

\path[draw=drawColor,line width= 0.2pt,line join=round] ( 29.60, 65.14) --
	( 92.80, 65.14);

\path[draw=drawColor,line width= 0.2pt,line join=round] ( 29.60, 84.36) --
	( 92.80, 84.36);

\path[draw=drawColor,line width= 0.2pt,line join=round] ( 32.47, 23.76) --
	( 32.47, 88.05);

\path[draw=drawColor,line width= 0.2pt,line join=round] ( 51.62, 23.76) --
	( 51.62, 88.05);

\path[draw=drawColor,line width= 0.2pt,line join=round] ( 70.78, 23.76) --
	( 70.78, 88.05);

\path[draw=drawColor,line width= 0.2pt,line join=round] ( 89.93, 23.76) --
	( 89.93, 88.05);
\definecolor{drawColor}{RGB}{50,162,81}

\path[draw=drawColor,line width= 0.9pt,line join=round] ( 32.47, 65.14) --
	( 37.26, 64.18) --
	( 42.05, 61.68) --
	( 46.83, 60.22) --
	( 51.62, 58.94) --
	( 56.41, 58.10) --
	( 61.20, 57.48) --
	( 65.99, 57.16) --
	( 70.78, 56.89) --
	( 75.57, 56.78) --
	( 80.35, 56.73) --
	( 85.14, 56.67) --
	( 89.93, 56.70);
\definecolor{drawColor}{RGB}{255,127,15}

\path[draw=drawColor,line width= 0.9pt,line join=round] ( 32.47, 65.14) --
	( 37.26, 63.37) --
	( 42.05, 60.41) --
	( 46.83, 59.09) --
	( 51.62, 58.24) --
	( 56.41, 57.79) --
	( 61.20, 57.49) --
	( 65.99, 57.55) --
	( 70.78, 57.63) --
	( 75.57, 57.68) --
	( 80.35, 57.80) --
	( 85.14, 57.98) --
	( 89.93, 58.03);
\definecolor{drawColor}{RGB}{60,183,204}

\path[draw=drawColor,line width= 0.9pt,line join=round] ( 32.47, 65.14) --
	( 37.26, 63.02) --
	( 42.05, 60.15) --
	( 46.83, 58.78) --
	( 51.62, 57.79) --
	( 56.41, 57.13) --
	( 61.20, 56.68) --
	( 65.99, 56.45) --
	( 70.78, 56.32) --
	( 75.57, 56.35) --
	( 80.35, 56.31) --
	( 85.14, 56.39) --
	( 89.93, 56.48);
\definecolor{drawColor}{gray}{0.50}

\path[draw=drawColor,line width= 0.6pt,line join=round,line cap=round] ( 29.60, 23.76) rectangle ( 92.80, 88.05);
\end{scope}
\begin{scope}
\path[clip] ( 92.80, 23.76) rectangle (156.01, 88.05);
\definecolor{fillColor}{RGB}{255,255,255}

\path[fill=fillColor] ( 92.80, 23.76) rectangle (156.01, 88.05);
\definecolor{drawColor}{gray}{0.98}

\path[draw=drawColor,line width= 0.6pt,line join=round] ( 92.80, 36.29) --
	(156.01, 36.29);

\path[draw=drawColor,line width= 0.6pt,line join=round] ( 92.80, 55.52) --
	(156.01, 55.52);

\path[draw=drawColor,line width= 0.6pt,line join=round] ( 92.80, 74.75) --
	(156.01, 74.75);

\path[draw=drawColor,line width= 0.6pt,line join=round] (105.25, 23.76) --
	(105.25, 88.05);

\path[draw=drawColor,line width= 0.6pt,line join=round] (124.41, 23.76) --
	(124.41, 88.05);

\path[draw=drawColor,line width= 0.6pt,line join=round] (143.56, 23.76) --
	(143.56, 88.05);
\definecolor{drawColor}{gray}{0.90}

\path[draw=drawColor,line width= 0.2pt,line join=round] ( 92.80, 26.68) --
	(156.01, 26.68);

\path[draw=drawColor,line width= 0.2pt,line join=round] ( 92.80, 45.91) --
	(156.01, 45.91);

\path[draw=drawColor,line width= 0.2pt,line join=round] ( 92.80, 65.14) --
	(156.01, 65.14);

\path[draw=drawColor,line width= 0.2pt,line join=round] ( 92.80, 84.36) --
	(156.01, 84.36);

\path[draw=drawColor,line width= 0.2pt,line join=round] ( 95.68, 23.76) --
	( 95.68, 88.05);

\path[draw=drawColor,line width= 0.2pt,line join=round] (114.83, 23.76) --
	(114.83, 88.05);

\path[draw=drawColor,line width= 0.2pt,line join=round] (133.98, 23.76) --
	(133.98, 88.05);

\path[draw=drawColor,line width= 0.2pt,line join=round] (153.14, 23.76) --
	(153.14, 88.05);
\definecolor{drawColor}{RGB}{50,162,81}

\path[draw=drawColor,line width= 0.9pt,line join=round] ( 95.68, 65.14) --
	(100.47, 65.33) --
	(105.25, 66.52) --
	(110.04, 66.61) --
	(114.83, 66.05) --
	(119.62, 65.24) --
	(124.41, 64.42) --
	(129.20, 63.53) --
	(133.98, 62.63) --
	(138.77, 61.73) --
	(143.56, 60.85) --
	(148.35, 60.01) --
	(153.14, 59.20);
\definecolor{drawColor}{RGB}{255,127,15}

\path[draw=drawColor,line width= 0.9pt,line join=round] ( 95.68, 65.14) --
	(100.47, 58.02) --
	(105.25, 54.29) --
	(110.04, 51.86) --
	(114.83, 50.21) --
	(119.62, 49.10) --
	(124.41, 48.30) --
	(129.20, 47.64) --
	(133.98, 47.23) --
	(138.77, 46.90) --
	(143.56, 46.63) --
	(148.35, 46.41) --
	(153.14, 46.24);
\definecolor{drawColor}{RGB}{60,183,204}

\path[draw=drawColor,line width= 0.9pt,line join=round] ( 95.68, 65.14) --
	(100.47, 57.93) --
	(105.25, 53.95) --
	(110.04, 51.74) --
	(114.83, 50.33) --
	(119.62, 49.50) --
	(124.41, 48.94) --
	(129.20, 48.50) --
	(133.98, 48.14) --
	(138.77, 47.90) --
	(143.56, 47.65) --
	(148.35, 47.41) --
	(153.14, 47.17);
\definecolor{drawColor}{gray}{0.50}

\path[draw=drawColor,line width= 0.6pt,line join=round,line cap=round] ( 92.80, 23.76) rectangle (156.01, 88.05);
\end{scope}
\begin{scope}
\path[clip] (156.01, 23.76) rectangle (219.22, 88.05);
\definecolor{fillColor}{RGB}{255,255,255}

\path[fill=fillColor] (156.01, 23.76) rectangle (219.22, 88.05);
\definecolor{drawColor}{gray}{0.98}

\path[draw=drawColor,line width= 0.6pt,line join=round] (156.01, 36.29) --
	(219.22, 36.29);

\path[draw=drawColor,line width= 0.6pt,line join=round] (156.01, 55.52) --
	(219.22, 55.52);

\path[draw=drawColor,line width= 0.6pt,line join=round] (156.01, 74.75) --
	(219.22, 74.75);

\path[draw=drawColor,line width= 0.6pt,line join=round] (168.46, 23.76) --
	(168.46, 88.05);

\path[draw=drawColor,line width= 0.6pt,line join=round] (187.62, 23.76) --
	(187.62, 88.05);

\path[draw=drawColor,line width= 0.6pt,line join=round] (206.77, 23.76) --
	(206.77, 88.05);
\definecolor{drawColor}{gray}{0.90}

\path[draw=drawColor,line width= 0.2pt,line join=round] (156.01, 26.68) --
	(219.22, 26.68);

\path[draw=drawColor,line width= 0.2pt,line join=round] (156.01, 45.91) --
	(219.22, 45.91);

\path[draw=drawColor,line width= 0.2pt,line join=round] (156.01, 65.14) --
	(219.22, 65.14);

\path[draw=drawColor,line width= 0.2pt,line join=round] (156.01, 84.36) --
	(219.22, 84.36);

\path[draw=drawColor,line width= 0.2pt,line join=round] (158.88, 23.76) --
	(158.88, 88.05);

\path[draw=drawColor,line width= 0.2pt,line join=round] (178.04, 23.76) --
	(178.04, 88.05);

\path[draw=drawColor,line width= 0.2pt,line join=round] (197.19, 23.76) --
	(197.19, 88.05);

\path[draw=drawColor,line width= 0.2pt,line join=round] (216.35, 23.76) --
	(216.35, 88.05);
\definecolor{drawColor}{RGB}{50,162,81}

\path[draw=drawColor,line width= 0.9pt,line join=round] (158.88, 65.14) --
	(163.67, 64.26) --
	(168.46, 63.00) --
	(173.25, 62.19) --
	(178.04, 61.50) --
	(182.83, 61.07) --
	(187.62, 60.60) --
	(192.40, 60.44) --
	(197.19, 60.13) --
	(201.98, 59.80) --
	(206.77, 59.74) --
	(211.56, 59.38) --
	(216.35, 59.27);
\definecolor{drawColor}{RGB}{255,127,15}

\path[draw=drawColor,line width= 0.9pt,line join=round] (158.88, 65.14) --
	(163.67, 64.01) --
	(168.46, 62.73) --
	(173.25, 62.01) --
	(178.04, 61.27) --
	(182.83, 60.92) --
	(187.62, 60.58) --
	(192.40, 60.58) --
	(197.19, 60.41) --
	(201.98, 60.25) --
	(206.77, 60.23) --
	(211.56, 59.91) --
	(216.35, 59.92);
\definecolor{drawColor}{RGB}{60,183,204}

\path[draw=drawColor,line width= 0.9pt,line join=round] (158.88, 65.14) --
	(163.67, 63.65) --
	(168.46, 62.37) --
	(173.25, 61.66) --
	(178.04, 61.01) --
	(182.83, 60.72) --
	(187.62, 60.36) --
	(192.40, 60.23) --
	(197.19, 59.98) --
	(201.98, 59.71) --
	(206.77, 59.66) --
	(211.56, 59.38) --
	(216.35, 59.33);
\definecolor{drawColor}{gray}{0.50}

\path[draw=drawColor,line width= 0.6pt,line join=round,line cap=round] (156.01, 23.76) rectangle (219.22, 88.05);
\end{scope}
\begin{scope}
\path[clip] ( 29.60, 88.05) rectangle ( 92.80, 97.56);
\definecolor{drawColor}{gray}{0.50}
\definecolor{fillColor}{gray}{0.80}

\path[draw=drawColor,line width= 0.2pt,line join=round,line cap=round,fill=fillColor] ( 29.60, 88.05) rectangle ( 92.80, 97.56);
\definecolor{drawColor}{gray}{0.10}

\node[text=drawColor,anchor=base,inner sep=0pt, outer sep=0pt, scale=  0.80] at ( 61.20, 90.05) {fluid height};
\end{scope}
\begin{scope}
\path[clip] ( 92.80, 88.05) rectangle (156.01, 97.56);
\definecolor{drawColor}{gray}{0.50}
\definecolor{fillColor}{gray}{0.80}

\path[draw=drawColor,line width= 0.2pt,line join=round,line cap=round,fill=fillColor] ( 92.80, 88.05) rectangle (156.01, 97.56);
\definecolor{drawColor}{gray}{0.10}

\node[text=drawColor,anchor=base,inner sep=0pt, outer sep=0pt, scale=  0.80] at (124.41, 90.05) {rain content};
\end{scope}
\begin{scope}
\path[clip] (156.01, 88.05) rectangle (219.22, 97.56);
\definecolor{drawColor}{gray}{0.50}
\definecolor{fillColor}{gray}{0.80}

\path[draw=drawColor,line width= 0.2pt,line join=round,line cap=round,fill=fillColor] (156.01, 88.05) rectangle (219.22, 97.56);
\definecolor{drawColor}{gray}{0.10}

\node[text=drawColor,anchor=base,inner sep=0pt, outer sep=0pt, scale=  0.80] at (187.62, 90.05) {wind};
\end{scope}
\begin{scope}
\path[clip] (  0.00,  0.00) rectangle (231.26, 97.56);
\definecolor{drawColor}{RGB}{0,0,0}

\node[text=drawColor,anchor=base east,inner sep=0pt, outer sep=0pt, scale=  0.72] at ( 24.20, 24.20) {50};

\node[text=drawColor,anchor=base east,inner sep=0pt, outer sep=0pt, scale=  0.72] at ( 24.20, 43.43) {75};

\node[text=drawColor,anchor=base east,inner sep=0pt, outer sep=0pt, scale=  0.72] at ( 24.20, 62.66) {100};

\node[text=drawColor,anchor=base east,inner sep=0pt, outer sep=0pt, scale=  0.72] at ( 24.20, 81.88) {125};
\end{scope}
\begin{scope}
\path[clip] (  0.00,  0.00) rectangle (231.26, 97.56);
\definecolor{drawColor}{RGB}{0,0,0}

\path[draw=drawColor,line width= 0.6pt,line join=round] ( 26.60, 26.68) --
	( 29.60, 26.68);

\path[draw=drawColor,line width= 0.6pt,line join=round] ( 26.60, 45.91) --
	( 29.60, 45.91);

\path[draw=drawColor,line width= 0.6pt,line join=round] ( 26.60, 65.14) --
	( 29.60, 65.14);

\path[draw=drawColor,line width= 0.6pt,line join=round] ( 26.60, 84.36) --
	( 29.60, 84.36);
\end{scope}
\begin{scope}
\path[clip] (  0.00,  0.00) rectangle (231.26, 97.56);
\definecolor{drawColor}{RGB}{0,0,0}

\path[draw=drawColor,line width= 0.6pt,line join=round] ( 32.47, 20.76) --
	( 32.47, 23.76);

\path[draw=drawColor,line width= 0.6pt,line join=round] ( 51.62, 20.76) --
	( 51.62, 23.76);

\path[draw=drawColor,line width= 0.6pt,line join=round] ( 70.78, 20.76) --
	( 70.78, 23.76);

\path[draw=drawColor,line width= 0.6pt,line join=round] ( 89.93, 20.76) --
	( 89.93, 23.76);
\end{scope}
\begin{scope}
\path[clip] (  0.00,  0.00) rectangle (231.26, 97.56);
\definecolor{drawColor}{RGB}{0,0,0}

\node[text=drawColor,anchor=base,inner sep=0pt, outer sep=0pt, scale=  0.72] at ( 32.47, 13.40) {0};

\node[text=drawColor,anchor=base,inner sep=0pt, outer sep=0pt, scale=  0.72] at ( 51.62, 13.40) {20};

\node[text=drawColor,anchor=base,inner sep=0pt, outer sep=0pt, scale=  0.72] at ( 70.78, 13.40) {40};

\node[text=drawColor,anchor=base,inner sep=0pt, outer sep=0pt, scale=  0.72] at ( 89.93, 13.40) {60};
\end{scope}
\begin{scope}
\path[clip] (  0.00,  0.00) rectangle (231.26, 97.56);
\definecolor{drawColor}{RGB}{0,0,0}

\path[draw=drawColor,line width= 0.6pt,line join=round] ( 95.68, 20.76) --
	( 95.68, 23.76);

\path[draw=drawColor,line width= 0.6pt,line join=round] (114.83, 20.76) --
	(114.83, 23.76);

\path[draw=drawColor,line width= 0.6pt,line join=round] (133.98, 20.76) --
	(133.98, 23.76);

\path[draw=drawColor,line width= 0.6pt,line join=round] (153.14, 20.76) --
	(153.14, 23.76);
\end{scope}
\begin{scope}
\path[clip] (  0.00,  0.00) rectangle (231.26, 97.56);
\definecolor{drawColor}{RGB}{0,0,0}

\node[text=drawColor,anchor=base,inner sep=0pt, outer sep=0pt, scale=  0.72] at ( 95.68, 13.40) {0};

\node[text=drawColor,anchor=base,inner sep=0pt, outer sep=0pt, scale=  0.72] at (114.83, 13.40) {20};

\node[text=drawColor,anchor=base,inner sep=0pt, outer sep=0pt, scale=  0.72] at (133.98, 13.40) {40};

\node[text=drawColor,anchor=base,inner sep=0pt, outer sep=0pt, scale=  0.72] at (153.14, 13.40) {60};
\end{scope}
\begin{scope}
\path[clip] (  0.00,  0.00) rectangle (231.26, 97.56);
\definecolor{drawColor}{RGB}{0,0,0}

\path[draw=drawColor,line width= 0.6pt,line join=round] (158.88, 20.76) --
	(158.88, 23.76);

\path[draw=drawColor,line width= 0.6pt,line join=round] (178.04, 20.76) --
	(178.04, 23.76);

\path[draw=drawColor,line width= 0.6pt,line join=round] (197.19, 20.76) --
	(197.19, 23.76);

\path[draw=drawColor,line width= 0.6pt,line join=round] (216.35, 20.76) --
	(216.35, 23.76);
\end{scope}
\begin{scope}
\path[clip] (  0.00,  0.00) rectangle (231.26, 97.56);
\definecolor{drawColor}{RGB}{0,0,0}

\node[text=drawColor,anchor=base,inner sep=0pt, outer sep=0pt, scale=  0.72] at (158.88, 13.40) {0};

\node[text=drawColor,anchor=base,inner sep=0pt, outer sep=0pt, scale=  0.72] at (178.04, 13.40) {20};

\node[text=drawColor,anchor=base,inner sep=0pt, outer sep=0pt, scale=  0.72] at (197.19, 13.40) {40};

\node[text=drawColor,anchor=base,inner sep=0pt, outer sep=0pt, scale=  0.72] at (216.35, 13.40) {60};
\end{scope}
\begin{scope}
\path[clip] (  0.00,  0.00) rectangle (231.26, 97.56);
\definecolor{drawColor}{RGB}{0,0,0}

\node[text=drawColor,anchor=base,inner sep=0pt, outer sep=0pt, scale=  0.90] at (124.41,  2.40) {time (min)};
\end{scope}
\begin{scope}
\path[clip] (  0.00,  0.00) rectangle (231.26, 97.56);
\definecolor{drawColor}{RGB}{0,0,0}

\node[text=drawColor,rotate= 90.00,anchor=base,inner sep=0pt, outer sep=0pt, scale=  0.90] at (  8.60, 55.91) {\textsc{crps}};
\end{scope}
\begin{scope}
\path[clip] (  0.00,  0.00) rectangle (231.26, 97.56);
\definecolor{drawColor}{RGB}{50,162,81}

\path[draw=drawColor,line width= 2.8pt,line join=round] (166.96, 80.50) -- (172.66, 80.50);
\end{scope}
\begin{scope}
\path[clip] (  0.00,  0.00) rectangle (231.26, 97.56);
\definecolor{drawColor}{RGB}{255,127,15}

\path[draw=drawColor,line width= 2.8pt,line join=round] (166.96, 73.39) -- (172.66, 73.39);
\end{scope}
\begin{scope}
\path[clip] (  0.00,  0.00) rectangle (231.26, 97.56);
\definecolor{drawColor}{RGB}{60,183,204}

\path[draw=drawColor,line width= 2.8pt,line join=round] (166.96, 66.28) -- (172.66, 66.28);
\end{scope}
\begin{scope}
\path[clip] (  0.00,  0.00) rectangle (231.26, 97.56);
\definecolor{drawColor}{RGB}{0,0,0}

\node[text=drawColor,anchor=base west,inner sep=0pt, outer sep=0pt, scale=  0.60] at (175.17, 78.44) {\textsc{le}n\textsc{kf}};
\end{scope}
\begin{scope}
\path[clip] (  0.00,  0.00) rectangle (231.26, 97.56);
\definecolor{drawColor}{RGB}{0,0,0}

\node[text=drawColor,anchor=base west,inner sep=0pt, outer sep=0pt, scale=  0.60] at (175.17, 71.32) {\textsc{naive-le}n\textsc{kpf}};
\end{scope}
\begin{scope}
\path[clip] (  0.00,  0.00) rectangle (231.26, 97.56);
\definecolor{drawColor}{RGB}{0,0,0}

\node[text=drawColor,anchor=base west,inner sep=0pt, outer sep=0pt, scale=  0.60] at (175.17, 64.21) {\textsc{block-le}n\textsc{kpf}};
\end{scope}
\end{tikzpicture}

%% file: HF_U-pattern.tex
\begin{tikzpicture}[x=1pt,y=1pt]
\definecolor{fillColor}{RGB}{255,255,255}
\path[use as bounding box,fill=fillColor,fill opacity=0.00] (0,0) rectangle (231.26,108.41);
\begin{scope}
\path[clip] ( 29.60, 23.76) rectangle (124.41, 98.90);
\definecolor{fillColor}{RGB}{255,255,255}

\path[fill=fillColor] ( 29.60, 23.76) rectangle (124.41, 98.90);
\definecolor{drawColor}{gray}{0.98}

\path[draw=drawColor,line width= 0.6pt,line join=round] ( 29.60, 38.41) --
	(124.41, 38.41);

\path[draw=drawColor,line width= 0.6pt,line join=round] ( 29.60, 60.88) --
	(124.41, 60.88);

\path[draw=drawColor,line width= 0.6pt,line join=round] ( 29.60, 83.35) --
	(124.41, 83.35);

\path[draw=drawColor,line width= 0.6pt,line join=round] ( 45.88, 23.76) --
	( 45.88, 98.90);

\path[draw=drawColor,line width= 0.6pt,line join=round] ( 69.82, 23.76) --
	( 69.82, 98.90);

\path[draw=drawColor,line width= 0.6pt,line join=round] ( 93.76, 23.76) --
	( 93.76, 98.90);

\path[draw=drawColor,line width= 0.6pt,line join=round] (117.70, 23.76) --
	(117.70, 98.90);
\definecolor{drawColor}{gray}{0.90}

\path[draw=drawColor,line width= 0.2pt,line join=round] ( 29.60, 27.17) --
	(124.41, 27.17);

\path[draw=drawColor,line width= 0.2pt,line join=round] ( 29.60, 49.64) --
	(124.41, 49.64);

\path[draw=drawColor,line width= 0.2pt,line join=round] ( 29.60, 72.11) --
	(124.41, 72.11);

\path[draw=drawColor,line width= 0.2pt,line join=round] ( 29.60, 94.58) --
	(124.41, 94.58);

\path[draw=drawColor,line width= 0.2pt,line join=round] ( 33.91, 23.76) --
	( 33.91, 98.90);

\path[draw=drawColor,line width= 0.2pt,line join=round] ( 57.85, 23.76) --
	( 57.85, 98.90);

\path[draw=drawColor,line width= 0.2pt,line join=round] ( 81.79, 23.76) --
	( 81.79, 98.90);

\path[draw=drawColor,line width= 0.2pt,line join=round] (105.73, 23.76) --
	(105.73, 98.90);
\definecolor{drawColor}{RGB}{50,162,81}

\path[draw=drawColor,line width= 0.9pt,line join=round] ( 33.91, 72.11) --
	( 35.10, 70.99) --
	( 36.30, 68.07) --
	( 37.50, 66.36) --
	( 38.69, 64.87) --
	( 39.89, 63.89) --
	( 41.09, 63.17) --
	( 42.29, 62.79) --
	( 43.48, 62.47) --
	( 44.68, 62.35) --
	( 45.88, 62.29) --
	( 47.07, 62.22) --
	( 48.27, 62.25) --
	( 49.47, 62.28) --
	( 50.67, 62.46) --
	( 51.86, 62.92) --
	( 53.06, 63.24) --
	( 54.26, 63.50) --
	( 55.45, 63.98) --
	( 56.65, 64.31) --
	( 57.85, 64.55) --
	( 59.04, 64.80) --
	( 60.24, 65.36) --
	( 61.44, 65.83) --
	( 62.64, 65.96) --
	( 63.83, 65.98) --
	( 65.03, 66.37) --
	( 66.23, 66.43) --
	( 67.42, 66.44) --
	( 68.62, 66.65) --
	( 69.82, 66.80) --
	( 71.02, 67.32) --
	( 72.21, 67.72) --
	( 73.41, 68.05) --
	( 74.61, 68.36) --
	( 75.80, 68.78) --
	( 77.00, 69.03) --
	( 78.20, 69.28) --
	( 79.40, 69.57) --
	( 80.59, 70.00) --
	( 81.79, 70.23) --
	( 82.99, 70.52) --
	( 84.18, 70.99) --
	( 85.38, 71.33) --
	( 86.58, 71.84) --
	( 87.78, 72.29) --
	( 88.97, 72.75) --
	( 90.17, 73.10) --
	( 91.37, 73.10) --
	( 92.56, 73.18) --
	( 93.76, 73.33) --
	( 94.96, 73.53) --
	( 96.16, 73.57) --
	( 97.35, 73.50) --
	( 98.55, 73.13) --
	( 99.75, 72.97) --
	(100.94, 72.94) --
	(102.14, 73.15) --
	(103.34, 73.31) --
	(104.54, 73.35) --
	(105.73, 73.41) --
	(106.93, 73.82) --
	(108.13, 74.01) --
	(109.32, 74.08) --
	(110.52, 74.25) --
	(111.72, 74.29) --
	(112.92, 74.69) --
	(114.11, 74.85) --
	(115.31, 74.87) --
	(116.51, 75.18) --
	(117.70, 75.47) --
	(118.90, 76.05) --
	(120.10, 76.28);
\definecolor{drawColor}{RGB}{255,127,15}

\path[draw=drawColor,line width= 0.9pt,line join=round] ( 33.91, 72.11) --
	( 35.10, 70.04) --
	( 36.30, 66.60) --
	( 37.50, 65.05) --
	( 38.69, 64.06) --
	( 39.89, 63.53) --
	( 41.09, 63.18) --
	( 42.29, 63.24) --
	( 43.48, 63.35) --
	( 44.68, 63.40) --
	( 45.88, 63.53) --
	( 47.07, 63.75) --
	( 48.27, 63.81) --
	( 49.47, 63.90) --
	( 50.67, 64.10) --
	( 51.86, 64.47) --
	( 53.06, 64.60) --
	( 54.26, 64.93) --
	( 55.45, 65.37) --
	( 56.65, 65.55) --
	( 57.85, 65.81) --
	( 59.04, 66.09) --
	( 60.24, 66.49) --
	( 61.44, 66.90) --
	( 62.64, 67.06) --
	( 63.83, 67.18) --
	( 65.03, 67.55) --
	( 66.23, 67.51) --
	( 67.42, 67.44) --
	( 68.62, 67.44) --
	( 69.82, 67.57) --
	( 71.02, 67.88) --
	( 72.21, 68.12) --
	( 73.41, 68.30) --
	( 74.61, 68.51) --
	( 75.80, 68.80) --
	( 77.00, 68.99) --
	( 78.20, 69.28) --
	( 79.40, 69.28) --
	( 80.59, 69.53) --
	( 81.79, 69.61) --
	( 82.99, 69.74) --
	( 84.18, 69.82) --
	( 85.38, 70.08) --
	( 86.58, 70.46) --
	( 87.78, 70.77) --
	( 88.97, 70.99) --
	( 90.17, 71.26) --
	( 91.37, 71.24) --
	( 92.56, 71.32) --
	( 93.76, 71.48) --
	( 94.96, 71.63) --
	( 96.16, 71.79) --
	( 97.35, 71.48) --
	( 98.55, 71.15) --
	( 99.75, 70.87) --
	(100.94, 70.69) --
	(102.14, 70.74) --
	(103.34, 70.86) --
	(104.54, 70.78) --
	(105.73, 70.90) --
	(106.93, 71.41) --
	(108.13, 71.48) --
	(109.32, 71.61) --
	(110.52, 71.73) --
	(111.72, 71.66) --
	(112.92, 71.93) --
	(114.11, 71.96) --
	(115.31, 71.92) --
	(116.51, 72.12) --
	(117.70, 72.33) --
	(118.90, 72.69) --
	(120.10, 72.77);
\definecolor{drawColor}{RGB}{60,183,204}

\path[draw=drawColor,line width= 0.9pt,line join=round] ( 33.91, 72.11) --
	( 35.10, 69.64) --
	( 36.30, 66.28) --
	( 37.50, 64.68) --
	( 38.69, 63.53) --
	( 39.89, 62.75) --
	( 41.09, 62.23) --
	( 42.29, 61.96) --
	( 43.48, 61.81) --
	( 44.68, 61.85) --
	( 45.88, 61.80) --
	( 47.07, 61.89) --
	( 48.27, 61.99) --
	( 49.47, 62.06) --
	( 50.67, 62.27) --
	( 51.86, 62.77) --
	( 53.06, 63.01) --
	( 54.26, 63.37) --
	( 55.45, 63.94) --
	( 56.65, 64.31) --
	( 57.85, 64.69) --
	( 59.04, 65.08) --
	( 60.24, 65.79) --
	( 61.44, 66.32) --
	( 62.64, 66.60) --
	( 63.83, 66.84) --
	( 65.03, 67.31) --
	( 66.23, 67.47) --
	( 67.42, 67.45) --
	( 68.62, 67.62) --
	( 69.82, 67.86) --
	( 71.02, 68.45) --
	( 72.21, 68.79) --
	( 73.41, 69.22) --
	( 74.61, 69.86) --
	( 75.80, 70.40) --
	( 77.00, 70.83) --
	( 78.20, 71.34) --
	( 79.40, 71.73) --
	( 80.59, 72.10) --
	( 81.79, 72.37) --
	( 82.99, 72.76) --
	( 84.18, 73.03) --
	( 85.38, 73.53) --
	( 86.58, 74.05) --
	( 87.78, 74.52) --
	( 88.97, 75.01) --
	( 90.17, 75.29) --
	( 91.37, 75.37) --
	( 92.56, 75.52) --
	( 93.76, 75.83) --
	( 94.96, 76.02) --
	( 96.16, 76.21) --
	( 97.35, 76.18) --
	( 98.55, 76.05) --
	( 99.75, 75.98) --
	(100.94, 76.07) --
	(102.14, 76.14) --
	(103.34, 76.23) --
	(104.54, 76.12) --
	(105.73, 76.30) --
	(106.93, 76.98) --
	(108.13, 77.26) --
	(109.32, 77.52) --
	(110.52, 77.80) --
	(111.72, 77.89) --
	(112.92, 78.22) --
	(114.11, 78.13) --
	(115.31, 78.03) --
	(116.51, 78.40) --
	(117.70, 78.73) --
	(118.90, 79.26) --
	(120.10, 79.34);
\definecolor{drawColor}{gray}{0.50}

\path[draw=drawColor,line width= 0.6pt,line join=round,line cap=round] ( 29.60, 23.76) rectangle (124.41, 98.90);
\end{scope}
\begin{scope}
\path[clip] (124.41, 23.76) rectangle (219.22, 98.90);
\definecolor{fillColor}{RGB}{255,255,255}

\path[fill=fillColor] (124.41, 23.76) rectangle (219.22, 98.90);
\definecolor{drawColor}{gray}{0.98}

\path[draw=drawColor,line width= 0.6pt,line join=round] (124.41, 38.41) --
	(219.22, 38.41);

\path[draw=drawColor,line width= 0.6pt,line join=round] (124.41, 60.88) --
	(219.22, 60.88);

\path[draw=drawColor,line width= 0.6pt,line join=round] (124.41, 83.35) --
	(219.22, 83.35);

\path[draw=drawColor,line width= 0.6pt,line join=round] (140.69, 23.76) --
	(140.69, 98.90);

\path[draw=drawColor,line width= 0.6pt,line join=round] (164.63, 23.76) --
	(164.63, 98.90);

\path[draw=drawColor,line width= 0.6pt,line join=round] (188.57, 23.76) --
	(188.57, 98.90);

\path[draw=drawColor,line width= 0.6pt,line join=round] (212.52, 23.76) --
	(212.52, 98.90);
\definecolor{drawColor}{gray}{0.90}

\path[draw=drawColor,line width= 0.2pt,line join=round] (124.41, 27.17) --
	(219.22, 27.17);

\path[draw=drawColor,line width= 0.2pt,line join=round] (124.41, 49.64) --
	(219.22, 49.64);

\path[draw=drawColor,line width= 0.2pt,line join=round] (124.41, 72.11) --
	(219.22, 72.11);

\path[draw=drawColor,line width= 0.2pt,line join=round] (124.41, 94.58) --
	(219.22, 94.58);

\path[draw=drawColor,line width= 0.2pt,line join=round] (128.72, 23.76) --
	(128.72, 98.90);

\path[draw=drawColor,line width= 0.2pt,line join=round] (152.66, 23.76) --
	(152.66, 98.90);

\path[draw=drawColor,line width= 0.2pt,line join=round] (176.60, 23.76) --
	(176.60, 98.90);

\path[draw=drawColor,line width= 0.2pt,line join=round] (200.54, 23.76) --
	(200.54, 98.90);
\definecolor{drawColor}{RGB}{50,162,81}

\path[draw=drawColor,line width= 0.9pt,line join=round] (128.72, 72.11) --
	(129.91, 72.34) --
	(131.11, 73.73) --
	(132.31, 73.83) --
	(133.51, 73.18) --
	(134.70, 72.24) --
	(135.90, 71.28) --
	(137.10, 70.23) --
	(138.29, 69.18) --
	(139.49, 68.14) --
	(140.69, 67.10) --
	(141.89, 66.12) --
	(143.08, 65.18) --
	(144.28, 64.22) --
	(145.48, 63.35) --
	(146.67, 62.56) --
	(147.87, 61.81) --
	(149.07, 61.17) --
	(150.27, 60.51) --
	(151.46, 59.84) --
	(152.66, 59.21) --
	(153.86, 58.60) --
	(155.05, 58.08) --
	(156.25, 57.60) --
	(157.45, 57.15) --
	(158.64, 56.65) --
	(159.84, 56.19) --
	(161.04, 55.76) --
	(162.24, 55.32) --
	(163.43, 54.90) --
	(164.63, 54.52) --
	(165.83, 54.09) --
	(167.02, 53.66) --
	(168.22, 53.26) --
	(169.42, 52.93) --
	(170.62, 52.59) --
	(171.81, 52.29) --
	(173.01, 51.97) --
	(174.21, 51.67) --
	(175.40, 51.33) --
	(176.60, 50.98) --
	(177.80, 50.72) --
	(179.00, 50.55) --
	(180.19, 50.43) --
	(181.39, 50.29) --
	(182.59, 50.10) --
	(183.78, 49.95) --
	(184.98, 49.83) --
	(186.18, 49.69) --
	(187.38, 49.54) --
	(188.57, 49.39) --
	(189.77, 49.31) --
	(190.97, 49.21) --
	(192.16, 49.12) --
	(193.36, 48.97) --
	(194.56, 48.86) --
	(195.76, 48.81) --
	(196.95, 48.68) --
	(198.15, 48.49) --
	(199.35, 48.31) --
	(200.54, 48.15) --
	(201.74, 48.01) --
	(202.94, 47.82) --
	(204.14, 47.61) --
	(205.33, 47.44) --
	(206.53, 47.24) --
	(207.73, 47.03) --
	(208.92, 46.86) --
	(210.12, 46.72) --
	(211.32, 46.60) --
	(212.52, 46.48) --
	(213.71, 46.34) --
	(214.91, 46.22);
\definecolor{drawColor}{RGB}{255,127,15}

\path[draw=drawColor,line width= 0.9pt,line join=round] (128.72, 72.11) --
	(129.91, 63.79) --
	(131.11, 59.44) --
	(132.31, 56.60) --
	(133.51, 54.68) --
	(134.70, 53.37) --
	(135.90, 52.44) --
	(137.10, 51.67) --
	(138.29, 51.18) --
	(139.49, 50.81) --
	(140.69, 50.49) --
	(141.89, 50.23) --
	(143.08, 50.03) --
	(144.28, 49.83) --
	(145.48, 49.72) --
	(146.67, 49.66) --
	(147.87, 49.56) --
	(149.07, 49.52) --
	(150.27, 49.48) --
	(151.46, 49.44) --
	(152.66, 49.45) --
	(153.86, 49.50) --
	(155.05, 49.53) --
	(156.25, 49.60) --
	(157.45, 49.61) --
	(158.64, 49.64) --
	(159.84, 49.68) --
	(161.04, 49.70) --
	(162.24, 49.67) --
	(163.43, 49.71) --
	(164.63, 49.77) --
	(165.83, 49.78) --
	(167.02, 49.79) --
	(168.22, 49.82) --
	(169.42, 49.86) --
	(170.62, 49.88) --
	(171.81, 49.94) --
	(173.01, 50.01) --
	(174.21, 50.03) --
	(175.40, 50.02) --
	(176.60, 49.99) --
	(177.80, 49.99) --
	(179.00, 50.02) --
	(180.19, 50.06) --
	(181.39, 50.10) --
	(182.59, 50.14) --
	(183.78, 50.19) --
	(184.98, 50.23) --
	(186.18, 50.25) --
	(187.38, 50.27) --
	(188.57, 50.33) --
	(189.77, 50.43) --
	(190.97, 50.55) --
	(192.16, 50.63) --
	(193.36, 50.73) --
	(194.56, 50.81) --
	(195.76, 50.87) --
	(196.95, 50.89) --
	(198.15, 50.89) --
	(199.35, 50.91) --
	(200.54, 50.92) --
	(201.74, 50.94) --
	(202.94, 50.91) --
	(204.14, 50.89) --
	(205.33, 50.92) --
	(206.53, 50.93) --
	(207.73, 50.93) --
	(208.92, 50.90) --
	(210.12, 50.89) --
	(211.32, 50.88) --
	(212.52, 50.82) --
	(213.71, 50.80) --
	(214.91, 50.80);
\definecolor{drawColor}{RGB}{60,183,204}

\path[draw=drawColor,line width= 0.9pt,line join=round] (128.72, 72.11) --
	(129.91, 63.69) --
	(131.11, 59.04) --
	(132.31, 56.46) --
	(133.51, 54.81) --
	(134.70, 53.84) --
	(135.90, 53.18) --
	(137.10, 52.67) --
	(138.29, 52.25) --
	(139.49, 51.98) --
	(140.69, 51.67) --
	(141.89, 51.39) --
	(143.08, 51.12) --
	(144.28, 50.82) --
	(145.48, 50.56) --
	(146.67, 50.35) --
	(147.87, 50.20) --
	(149.07, 50.03) --
	(150.27, 49.89) --
	(151.46, 49.74) --
	(152.66, 49.52) --
	(153.86, 49.38) --
	(155.05, 49.26) --
	(156.25, 49.14) --
	(157.45, 49.01) --
	(158.64, 48.90) --
	(159.84, 48.83) --
	(161.04, 48.76) --
	(162.24, 48.68) --
	(163.43, 48.56) --
	(164.63, 48.49) --
	(165.83, 48.37) --
	(167.02, 48.26) --
	(168.22, 48.13) --
	(169.42, 48.07) --
	(170.62, 47.99) --
	(171.81, 47.96) --
	(173.01, 47.96) --
	(174.21, 47.92) --
	(175.40, 47.86) --
	(176.60, 47.77) --
	(177.80, 47.74) --
	(179.00, 47.75) --
	(180.19, 47.75) --
	(181.39, 47.77) --
	(182.59, 47.79) --
	(183.78, 47.79) --
	(184.98, 47.81) --
	(186.18, 47.85) --
	(187.38, 47.87) --
	(188.57, 47.96) --
	(189.77, 48.09) --
	(190.97, 48.18) --
	(192.16, 48.25) --
	(193.36, 48.31) --
	(194.56, 48.36) --
	(195.76, 48.45) --
	(196.95, 48.46) --
	(198.15, 48.41) --
	(199.35, 48.40) --
	(200.54, 48.38) --
	(201.74, 48.37) --
	(202.94, 48.35) --
	(204.14, 48.32) --
	(205.33, 48.32) --
	(206.53, 48.30) --
	(207.73, 48.28) --
	(208.92, 48.27) --
	(210.12, 48.28) --
	(211.32, 48.30) --
	(212.52, 48.30) --
	(213.71, 48.31) --
	(214.91, 48.31);
\definecolor{drawColor}{gray}{0.50}

\path[draw=drawColor,line width= 0.6pt,line join=round,line cap=round] (124.41, 23.76) rectangle (219.22, 98.90);
\end{scope}
\begin{scope}
\path[clip] ( 29.60, 98.90) rectangle (124.41,108.41);
\definecolor{drawColor}{gray}{0.50}
\definecolor{fillColor}{gray}{0.80}

\path[draw=drawColor,line width= 0.2pt,line join=round,line cap=round,fill=fillColor] ( 29.60, 98.90) rectangle (124.41,108.41);
\definecolor{drawColor}{gray}{0.10}

\node[text=drawColor,anchor=base,inner sep=0pt, outer sep=0pt, scale=  0.80] at ( 77.00,100.90) {fluid height};
\end{scope}
\begin{scope}
\path[clip] (124.41, 98.90) rectangle (219.22,108.41);
\definecolor{drawColor}{gray}{0.50}
\definecolor{fillColor}{gray}{0.80}

\path[draw=drawColor,line width= 0.2pt,line join=round,line cap=round,fill=fillColor] (124.41, 98.90) rectangle (219.22,108.41);
\definecolor{drawColor}{gray}{0.10}

\node[text=drawColor,anchor=base,inner sep=0pt, outer sep=0pt, scale=  0.80] at (171.81,100.90) {rain content};
\end{scope}
\begin{scope}
\path[clip] (  0.00,  0.00) rectangle (231.26,108.41);
\definecolor{drawColor}{RGB}{0,0,0}

\node[text=drawColor,anchor=base east,inner sep=0pt, outer sep=0pt, scale=  0.72] at ( 24.20, 24.69) {50};

\node[text=drawColor,anchor=base east,inner sep=0pt, outer sep=0pt, scale=  0.72] at ( 24.20, 47.16) {75};

\node[text=drawColor,anchor=base east,inner sep=0pt, outer sep=0pt, scale=  0.72] at ( 24.20, 69.63) {100};

\node[text=drawColor,anchor=base east,inner sep=0pt, outer sep=0pt, scale=  0.72] at ( 24.20, 92.10) {125};
\end{scope}
\begin{scope}
\path[clip] (  0.00,  0.00) rectangle (231.26,108.41);
\definecolor{drawColor}{RGB}{0,0,0}

\path[draw=drawColor,line width= 0.6pt,line join=round] ( 26.60, 27.17) --
	( 29.60, 27.17);

\path[draw=drawColor,line width= 0.6pt,line join=round] ( 26.60, 49.64) --
	( 29.60, 49.64);

\path[draw=drawColor,line width= 0.6pt,line join=round] ( 26.60, 72.11) --
	( 29.60, 72.11);

\path[draw=drawColor,line width= 0.6pt,line join=round] ( 26.60, 94.58) --
	( 29.60, 94.58);
\end{scope}
\begin{scope}
\path[clip] (  0.00,  0.00) rectangle (231.26,108.41);
\definecolor{drawColor}{RGB}{0,0,0}

\path[draw=drawColor,line width= 0.6pt,line join=round] ( 33.91, 20.76) --
	( 33.91, 23.76);

\path[draw=drawColor,line width= 0.6pt,line join=round] ( 57.85, 20.76) --
	( 57.85, 23.76);

\path[draw=drawColor,line width= 0.6pt,line join=round] ( 81.79, 20.76) --
	( 81.79, 23.76);

\path[draw=drawColor,line width= 0.6pt,line join=round] (105.73, 20.76) --
	(105.73, 23.76);
\end{scope}
\begin{scope}
\path[clip] (  0.00,  0.00) rectangle (231.26,108.41);
\definecolor{drawColor}{RGB}{0,0,0}

\node[text=drawColor,anchor=base,inner sep=0pt, outer sep=0pt, scale=  0.72] at ( 33.91, 13.40) {0};

\node[text=drawColor,anchor=base,inner sep=0pt, outer sep=0pt, scale=  0.72] at ( 57.85, 13.40) {100};

\node[text=drawColor,anchor=base,inner sep=0pt, outer sep=0pt, scale=  0.72] at ( 81.79, 13.40) {200};

\node[text=drawColor,anchor=base,inner sep=0pt, outer sep=0pt, scale=  0.72] at (105.73, 13.40) {300};
\end{scope}
\begin{scope}
\path[clip] (  0.00,  0.00) rectangle (231.26,108.41);
\definecolor{drawColor}{RGB}{0,0,0}

\path[draw=drawColor,line width= 0.6pt,line join=round] (128.72, 20.76) --
	(128.72, 23.76);

\path[draw=drawColor,line width= 0.6pt,line join=round] (152.66, 20.76) --
	(152.66, 23.76);

\path[draw=drawColor,line width= 0.6pt,line join=round] (176.60, 20.76) --
	(176.60, 23.76);

\path[draw=drawColor,line width= 0.6pt,line join=round] (200.54, 20.76) --
	(200.54, 23.76);
\end{scope}
\begin{scope}
\path[clip] (  0.00,  0.00) rectangle (231.26,108.41);
\definecolor{drawColor}{RGB}{0,0,0}

\node[text=drawColor,anchor=base,inner sep=0pt, outer sep=0pt, scale=  0.72] at (128.72, 13.40) {0};

\node[text=drawColor,anchor=base,inner sep=0pt, outer sep=0pt, scale=  0.72] at (152.66, 13.40) {100};

\node[text=drawColor,anchor=base,inner sep=0pt, outer sep=0pt, scale=  0.72] at (176.60, 13.40) {200};

\node[text=drawColor,anchor=base,inner sep=0pt, outer sep=0pt, scale=  0.72] at (200.54, 13.40) {300};
\end{scope}
\begin{scope}
\path[clip] (  0.00,  0.00) rectangle (231.26,108.41);
\definecolor{drawColor}{RGB}{0,0,0}

\node[text=drawColor,anchor=base,inner sep=0pt, outer sep=0pt, scale=  0.90] at (124.41,  2.40) {time (min)};
\end{scope}
\begin{scope}
\path[clip] (  0.00,  0.00) rectangle (231.26,108.41);
\definecolor{drawColor}{RGB}{0,0,0}

\node[text=drawColor,rotate= 90.00,anchor=base,inner sep=0pt, outer sep=0pt, scale=  0.90] at (  8.60, 61.33) {\textsc{crps}};
\end{scope}
\begin{scope}
\path[clip] (  0.00,  0.00) rectangle (231.26,108.41);
\definecolor{drawColor}{RGB}{50,162,81}

\path[draw=drawColor,line width= 2.8pt,line join=round] ( 62.67, 51.61) -- ( 68.36, 51.61);
\end{scope}
\begin{scope}
\path[clip] (  0.00,  0.00) rectangle (231.26,108.41);
\definecolor{drawColor}{RGB}{255,127,15}

\path[draw=drawColor,line width= 2.8pt,line join=round] ( 62.67, 44.49) -- ( 68.36, 44.49);
\end{scope}
\begin{scope}
\path[clip] (  0.00,  0.00) rectangle (231.26,108.41);
\definecolor{drawColor}{RGB}{60,183,204}

\path[draw=drawColor,line width= 2.8pt,line join=round] ( 62.67, 37.38) -- ( 68.36, 37.38);
\end{scope}
\begin{scope}
\path[clip] (  0.00,  0.00) rectangle (231.26,108.41);
\definecolor{drawColor}{RGB}{0,0,0}

\node[text=drawColor,anchor=base west,inner sep=0pt, outer sep=0pt, scale=  0.60] at ( 70.88, 49.54) {\textsc{le}n\textsc{kf}};
\end{scope}
\begin{scope}
\path[clip] (  0.00,  0.00) rectangle (231.26,108.41);
\definecolor{drawColor}{RGB}{0,0,0}

\node[text=drawColor,anchor=base west,inner sep=0pt, outer sep=0pt, scale=  0.60] at ( 70.88, 42.43) {\textsc{naive-le}n\textsc{kpf}};
\end{scope}
\begin{scope}
\path[clip] (  0.00,  0.00) rectangle (231.26,108.41);
\definecolor{drawColor}{RGB}{0,0,0}

\node[text=drawColor,anchor=base west,inner sep=0pt, outer sep=0pt, scale=  0.60] at ( 70.88, 35.31) {\textsc{block-le}n\textsc{kpf}};
\end{scope}
\end{tikzpicture}

%% file: calibration.tex
\begin{tikzpicture}[x=1pt,y=1pt]
\definecolor{fillColor}{RGB}{255,255,255}
\path[use as bounding box,fill=fillColor,fill opacity=0.00] (0,0) rectangle (231.26, 86.72);
\begin{scope}
\path[clip] (  0.00,  0.00) rectangle (231.26, 86.72);
\definecolor{drawColor}{RGB}{255,255,255}
\definecolor{fillColor}{RGB}{255,255,255}

\path[draw=drawColor,line width= 0.6pt,line join=round,line cap=round,fill=fillColor] (  0.00,  0.00) rectangle (231.26, 86.72);
\end{scope}
\begin{scope}
\path[clip] ( 31.60, 23.76) rectangle ( 94.14, 77.21);
\definecolor{fillColor}{RGB}{255,255,255}

\path[fill=fillColor] ( 31.60, 23.76) rectangle ( 94.14, 77.21);
\definecolor{drawColor}{gray}{0.98}

\path[draw=drawColor,line width= 0.6pt,line join=round] ( 31.60, 30.90) --
	( 94.14, 30.90);

\path[draw=drawColor,line width= 0.6pt,line join=round] ( 31.60, 40.34) --
	( 94.14, 40.34);

\path[draw=drawColor,line width= 0.6pt,line join=round] ( 31.60, 49.77) --
	( 94.14, 49.77);

\path[draw=drawColor,line width= 0.6pt,line join=round] ( 31.60, 59.21) --
	( 94.14, 59.21);

\path[draw=drawColor,line width= 0.6pt,line join=round] ( 31.60, 68.64) --
	( 94.14, 68.64);

\path[draw=drawColor,line width= 0.6pt,line join=round] ( 37.11, 23.76) --
	( 37.11, 77.21);

\path[draw=drawColor,line width= 0.6pt,line join=round] ( 49.38, 23.76) --
	( 49.38, 77.21);

\path[draw=drawColor,line width= 0.6pt,line join=round] ( 61.64, 23.76) --
	( 61.64, 77.21);

\path[draw=drawColor,line width= 0.6pt,line join=round] ( 73.90, 23.76) --
	( 73.90, 77.21);

\path[draw=drawColor,line width= 0.6pt,line join=round] ( 86.17, 23.76) --
	( 86.17, 77.21);
\definecolor{drawColor}{gray}{0.90}

\path[draw=drawColor,line width= 0.2pt,line join=round] ( 31.60, 26.19) --
	( 94.14, 26.19);

\path[draw=drawColor,line width= 0.2pt,line join=round] ( 31.60, 35.62) --
	( 94.14, 35.62);

\path[draw=drawColor,line width= 0.2pt,line join=round] ( 31.60, 45.06) --
	( 94.14, 45.06);

\path[draw=drawColor,line width= 0.2pt,line join=round] ( 31.60, 54.49) --
	( 94.14, 54.49);

\path[draw=drawColor,line width= 0.2pt,line join=round] ( 31.60, 63.92) --
	( 94.14, 63.92);

\path[draw=drawColor,line width= 0.2pt,line join=round] ( 31.60, 73.36) --
	( 94.14, 73.36);

\path[draw=drawColor,line width= 0.2pt,line join=round] ( 43.25, 23.76) --
	( 43.25, 77.21);

\path[draw=drawColor,line width= 0.2pt,line join=round] ( 55.51, 23.76) --
	( 55.51, 77.21);

\path[draw=drawColor,line width= 0.2pt,line join=round] ( 67.77, 23.76) --
	( 67.77, 77.21);

\path[draw=drawColor,line width= 0.2pt,line join=round] ( 80.03, 23.76) --
	( 80.03, 77.21);

\path[draw=drawColor,line width= 0.2pt,line join=round] ( 92.30, 23.76) --
	( 92.30, 77.21);
\definecolor{fillColor}{gray}{0.35}

\path[fill=fillColor] ( 31.60, 26.19) rectangle ( 32.82, 50.46);

\path[fill=fillColor] ( 32.82, 26.19) rectangle ( 34.05, 45.08);

\path[fill=fillColor] ( 34.05, 26.19) rectangle ( 35.27, 45.65);

\path[fill=fillColor] ( 35.27, 26.19) rectangle ( 36.50, 44.07);

\path[fill=fillColor] ( 36.50, 26.19) rectangle ( 37.73, 44.25);

\path[fill=fillColor] ( 37.73, 26.19) rectangle ( 38.95, 43.83);

\path[fill=fillColor] ( 38.95, 26.19) rectangle ( 40.18, 43.97);

\path[fill=fillColor] ( 40.18, 26.19) rectangle ( 41.41, 42.89);

\path[fill=fillColor] ( 41.41, 26.19) rectangle ( 42.63, 43.34);

\path[fill=fillColor] ( 42.63, 26.19) rectangle ( 43.86, 43.36);

\path[fill=fillColor] ( 43.86, 26.19) rectangle ( 45.08, 43.28);

\path[fill=fillColor] ( 45.08, 26.19) rectangle ( 46.31, 43.97);

\path[fill=fillColor] ( 46.31, 26.19) rectangle ( 47.54, 43.22);

\path[fill=fillColor] ( 47.54, 26.19) rectangle ( 48.76, 42.16);

\path[fill=fillColor] ( 48.76, 26.19) rectangle ( 49.99, 43.32);

\path[fill=fillColor] ( 49.99, 26.19) rectangle ( 51.22, 42.66);

\path[fill=fillColor] ( 51.22, 26.19) rectangle ( 52.44, 43.09);

\path[fill=fillColor] ( 52.44, 26.19) rectangle ( 53.67, 42.24);

\path[fill=fillColor] ( 53.67, 26.19) rectangle ( 54.90, 43.73);

\path[fill=fillColor] ( 54.90, 26.19) rectangle ( 56.12, 43.13);

\path[fill=fillColor] ( 56.12, 26.19) rectangle ( 57.35, 42.64);

\path[fill=fillColor] ( 57.35, 26.19) rectangle ( 58.57, 42.99);

\path[fill=fillColor] ( 58.57, 26.19) rectangle ( 59.80, 43.18);

\path[fill=fillColor] ( 59.80, 26.19) rectangle ( 61.03, 43.63);

\path[fill=fillColor] ( 61.03, 26.19) rectangle ( 62.25, 43.24);

\path[fill=fillColor] ( 62.25, 26.19) rectangle ( 63.48, 42.39);

\path[fill=fillColor] ( 63.48, 26.19) rectangle ( 64.71, 43.39);

\path[fill=fillColor] ( 64.71, 26.19) rectangle ( 65.93, 43.47);

\path[fill=fillColor] ( 65.93, 26.19) rectangle ( 67.16, 43.15);

\path[fill=fillColor] ( 67.16, 26.19) rectangle ( 68.38, 43.51);

\path[fill=fillColor] ( 68.38, 26.19) rectangle ( 69.61, 43.54);

\path[fill=fillColor] ( 69.61, 26.19) rectangle ( 70.84, 43.31);

\path[fill=fillColor] ( 70.84, 26.19) rectangle ( 72.06, 44.02);

\path[fill=fillColor] ( 72.06, 26.19) rectangle ( 73.29, 44.83);

\path[fill=fillColor] ( 73.29, 26.19) rectangle ( 74.52, 44.43);

\path[fill=fillColor] ( 74.52, 26.19) rectangle ( 75.74, 44.33);

\path[fill=fillColor] ( 75.74, 26.19) rectangle ( 76.97, 45.73);

\path[fill=fillColor] ( 76.97, 26.19) rectangle ( 78.19, 44.08);

\path[fill=fillColor] ( 78.19, 26.19) rectangle ( 79.42, 44.65);

\path[fill=fillColor] ( 79.42, 26.19) rectangle ( 80.65, 44.90);

\path[fill=fillColor] ( 80.65, 26.19) rectangle ( 81.87, 45.99);

\path[fill=fillColor] ( 81.87, 26.19) rectangle ( 83.10, 45.61);

\path[fill=fillColor] ( 83.10, 26.19) rectangle ( 84.33, 45.16);

\path[fill=fillColor] ( 84.33, 26.19) rectangle ( 85.55, 46.72);

\path[fill=fillColor] ( 85.55, 26.19) rectangle ( 86.78, 45.16);

\path[fill=fillColor] ( 86.78, 26.19) rectangle ( 88.01, 46.77);

\path[fill=fillColor] ( 88.01, 26.19) rectangle ( 89.23, 46.23);

\path[fill=fillColor] ( 89.23, 26.19) rectangle ( 90.46, 46.23);

\path[fill=fillColor] ( 90.46, 26.19) rectangle ( 91.68, 47.54);

\path[fill=fillColor] ( 91.68, 26.19) rectangle ( 92.91, 47.93);

\path[fill=fillColor] ( 92.91, 26.19) rectangle ( 94.14, 62.55);
\definecolor{drawColor}{gray}{0.50}

\path[draw=drawColor,line width= 0.6pt,line join=round,line cap=round] ( 31.60, 23.76) rectangle ( 94.14, 77.21);
\end{scope}
\begin{scope}
\path[clip] ( 94.14, 23.76) rectangle (156.68, 77.21);
\definecolor{fillColor}{RGB}{255,255,255}

\path[fill=fillColor] ( 94.14, 23.76) rectangle (156.68, 77.21);
\definecolor{drawColor}{gray}{0.98}

\path[draw=drawColor,line width= 0.6pt,line join=round] ( 94.14, 30.90) --
	(156.68, 30.90);

\path[draw=drawColor,line width= 0.6pt,line join=round] ( 94.14, 40.34) --
	(156.68, 40.34);

\path[draw=drawColor,line width= 0.6pt,line join=round] ( 94.14, 49.77) --
	(156.68, 49.77);

\path[draw=drawColor,line width= 0.6pt,line join=round] ( 94.14, 59.21) --
	(156.68, 59.21);

\path[draw=drawColor,line width= 0.6pt,line join=round] ( 94.14, 68.64) --
	(156.68, 68.64);

\path[draw=drawColor,line width= 0.6pt,line join=round] ( 99.65, 23.76) --
	( 99.65, 77.21);

\path[draw=drawColor,line width= 0.6pt,line join=round] (111.92, 23.76) --
	(111.92, 77.21);

\path[draw=drawColor,line width= 0.6pt,line join=round] (124.18, 23.76) --
	(124.18, 77.21);

\path[draw=drawColor,line width= 0.6pt,line join=round] (136.44, 23.76) --
	(136.44, 77.21);

\path[draw=drawColor,line width= 0.6pt,line join=round] (148.71, 23.76) --
	(148.71, 77.21);
\definecolor{drawColor}{gray}{0.90}

\path[draw=drawColor,line width= 0.2pt,line join=round] ( 94.14, 26.19) --
	(156.68, 26.19);

\path[draw=drawColor,line width= 0.2pt,line join=round] ( 94.14, 35.62) --
	(156.68, 35.62);

\path[draw=drawColor,line width= 0.2pt,line join=round] ( 94.14, 45.06) --
	(156.68, 45.06);

\path[draw=drawColor,line width= 0.2pt,line join=round] ( 94.14, 54.49) --
	(156.68, 54.49);

\path[draw=drawColor,line width= 0.2pt,line join=round] ( 94.14, 63.92) --
	(156.68, 63.92);

\path[draw=drawColor,line width= 0.2pt,line join=round] ( 94.14, 73.36) --
	(156.68, 73.36);

\path[draw=drawColor,line width= 0.2pt,line join=round] (105.79, 23.76) --
	(105.79, 77.21);

\path[draw=drawColor,line width= 0.2pt,line join=round] (118.05, 23.76) --
	(118.05, 77.21);

\path[draw=drawColor,line width= 0.2pt,line join=round] (130.31, 23.76) --
	(130.31, 77.21);

\path[draw=drawColor,line width= 0.2pt,line join=round] (142.58, 23.76) --
	(142.58, 77.21);

\path[draw=drawColor,line width= 0.2pt,line join=round] (154.84, 23.76) --
	(154.84, 77.21);
\definecolor{fillColor}{gray}{0.35}

\path[fill=fillColor] ( 94.14, 26.19) rectangle ( 95.36, 46.64);

\path[fill=fillColor] ( 95.36, 26.19) rectangle ( 96.59, 45.13);

\path[fill=fillColor] ( 96.59, 26.19) rectangle ( 97.82, 44.35);

\path[fill=fillColor] ( 97.82, 26.19) rectangle ( 99.04, 44.38);

\path[fill=fillColor] ( 99.04, 26.19) rectangle (100.27, 44.28);

\path[fill=fillColor] (100.27, 26.19) rectangle (101.49, 43.60);

\path[fill=fillColor] (101.49, 26.19) rectangle (102.72, 44.47);

\path[fill=fillColor] (102.72, 26.19) rectangle (103.95, 44.62);

\path[fill=fillColor] (103.95, 26.19) rectangle (105.17, 44.38);

\path[fill=fillColor] (105.17, 26.19) rectangle (106.40, 44.08);

\path[fill=fillColor] (106.40, 26.19) rectangle (107.63, 44.97);

\path[fill=fillColor] (107.63, 26.19) rectangle (108.85, 44.26);

\path[fill=fillColor] (108.85, 26.19) rectangle (110.08, 44.34);

\path[fill=fillColor] (110.08, 26.19) rectangle (111.30, 43.53);

\path[fill=fillColor] (111.30, 26.19) rectangle (112.53, 44.25);

\path[fill=fillColor] (112.53, 26.19) rectangle (113.76, 43.38);

\path[fill=fillColor] (113.76, 26.19) rectangle (114.98, 43.77);

\path[fill=fillColor] (114.98, 26.19) rectangle (116.21, 43.93);

\path[fill=fillColor] (116.21, 26.19) rectangle (117.44, 44.26);

\path[fill=fillColor] (117.44, 26.19) rectangle (118.66, 43.50);

\path[fill=fillColor] (118.66, 26.19) rectangle (119.89, 43.54);

\path[fill=fillColor] (119.89, 26.19) rectangle (121.12, 44.05);

\path[fill=fillColor] (121.12, 26.19) rectangle (122.34, 44.36);

\path[fill=fillColor] (122.34, 26.19) rectangle (123.57, 45.09);

\path[fill=fillColor] (123.57, 26.19) rectangle (124.79, 44.65);

\path[fill=fillColor] (124.79, 26.19) rectangle (126.02, 44.09);

\path[fill=fillColor] (126.02, 26.19) rectangle (127.25, 44.34);

\path[fill=fillColor] (127.25, 26.19) rectangle (128.47, 43.85);

\path[fill=fillColor] (128.47, 26.19) rectangle (129.70, 44.01);

\path[fill=fillColor] (129.70, 26.19) rectangle (130.93, 43.41);

\path[fill=fillColor] (130.93, 26.19) rectangle (132.15, 44.15);

\path[fill=fillColor] (132.15, 26.19) rectangle (133.38, 44.67);

\path[fill=fillColor] (133.38, 26.19) rectangle (134.60, 44.38);

\path[fill=fillColor] (134.60, 26.19) rectangle (135.83, 43.36);

\path[fill=fillColor] (135.83, 26.19) rectangle (137.06, 44.10);

\path[fill=fillColor] (137.06, 26.19) rectangle (138.28, 44.14);

\path[fill=fillColor] (138.28, 26.19) rectangle (139.51, 44.65);

\path[fill=fillColor] (139.51, 26.19) rectangle (140.74, 44.51);

\path[fill=fillColor] (140.74, 26.19) rectangle (141.96, 44.23);

\path[fill=fillColor] (141.96, 26.19) rectangle (143.19, 43.68);

\path[fill=fillColor] (143.19, 26.19) rectangle (144.41, 44.01);

\path[fill=fillColor] (144.41, 26.19) rectangle (145.64, 43.35);

\path[fill=fillColor] (145.64, 26.19) rectangle (146.87, 43.13);

\path[fill=fillColor] (146.87, 26.19) rectangle (148.09, 44.33);

\path[fill=fillColor] (148.09, 26.19) rectangle (149.32, 43.37);

\path[fill=fillColor] (149.32, 26.19) rectangle (150.55, 43.61);

\path[fill=fillColor] (150.55, 26.19) rectangle (151.77, 43.92);

\path[fill=fillColor] (151.77, 26.19) rectangle (153.00, 42.86);

\path[fill=fillColor] (153.00, 26.19) rectangle (154.23, 43.31);

\path[fill=fillColor] (154.23, 26.19) rectangle (155.45, 42.92);

\path[fill=fillColor] (155.45, 26.19) rectangle (156.68, 74.78);
\definecolor{drawColor}{gray}{0.50}

\path[draw=drawColor,line width= 0.6pt,line join=round,line cap=round] ( 94.14, 23.76) rectangle (156.68, 77.21);
\end{scope}
\begin{scope}
\path[clip] (156.68, 23.76) rectangle (219.22, 77.21);
\definecolor{fillColor}{RGB}{255,255,255}

\path[fill=fillColor] (156.68, 23.76) rectangle (219.22, 77.21);
\definecolor{drawColor}{gray}{0.98}

\path[draw=drawColor,line width= 0.6pt,line join=round] (156.68, 30.90) --
	(219.22, 30.90);

\path[draw=drawColor,line width= 0.6pt,line join=round] (156.68, 40.34) --
	(219.22, 40.34);

\path[draw=drawColor,line width= 0.6pt,line join=round] (156.68, 49.77) --
	(219.22, 49.77);

\path[draw=drawColor,line width= 0.6pt,line join=round] (156.68, 59.21) --
	(219.22, 59.21);

\path[draw=drawColor,line width= 0.6pt,line join=round] (156.68, 68.64) --
	(219.22, 68.64);

\path[draw=drawColor,line width= 0.6pt,line join=round] (162.20, 23.76) --
	(162.20, 77.21);

\path[draw=drawColor,line width= 0.6pt,line join=round] (174.46, 23.76) --
	(174.46, 77.21);

\path[draw=drawColor,line width= 0.6pt,line join=round] (186.72, 23.76) --
	(186.72, 77.21);

\path[draw=drawColor,line width= 0.6pt,line join=round] (198.99, 23.76) --
	(198.99, 77.21);

\path[draw=drawColor,line width= 0.6pt,line join=round] (211.25, 23.76) --
	(211.25, 77.21);
\definecolor{drawColor}{gray}{0.90}

\path[draw=drawColor,line width= 0.2pt,line join=round] (156.68, 26.19) --
	(219.22, 26.19);

\path[draw=drawColor,line width= 0.2pt,line join=round] (156.68, 35.62) --
	(219.22, 35.62);

\path[draw=drawColor,line width= 0.2pt,line join=round] (156.68, 45.06) --
	(219.22, 45.06);

\path[draw=drawColor,line width= 0.2pt,line join=round] (156.68, 54.49) --
	(219.22, 54.49);

\path[draw=drawColor,line width= 0.2pt,line join=round] (156.68, 63.92) --
	(219.22, 63.92);

\path[draw=drawColor,line width= 0.2pt,line join=round] (156.68, 73.36) --
	(219.22, 73.36);

\path[draw=drawColor,line width= 0.2pt,line join=round] (168.33, 23.76) --
	(168.33, 77.21);

\path[draw=drawColor,line width= 0.2pt,line join=round] (180.59, 23.76) --
	(180.59, 77.21);

\path[draw=drawColor,line width= 0.2pt,line join=round] (192.85, 23.76) --
	(192.85, 77.21);

\path[draw=drawColor,line width= 0.2pt,line join=round] (205.12, 23.76) --
	(205.12, 77.21);

\path[draw=drawColor,line width= 0.2pt,line join=round] (217.38, 23.76) --
	(217.38, 77.21);
\definecolor{fillColor}{gray}{0.35}

\path[fill=fillColor] (156.68, 26.19) rectangle (157.90, 47.00);

\path[fill=fillColor] (157.90, 26.19) rectangle (159.13, 45.77);

\path[fill=fillColor] (159.13, 26.19) rectangle (160.36, 45.82);

\path[fill=fillColor] (160.36, 26.19) rectangle (161.58, 45.33);

\path[fill=fillColor] (161.58, 26.19) rectangle (162.81, 45.30);

\path[fill=fillColor] (162.81, 26.19) rectangle (164.04, 45.45);

\path[fill=fillColor] (164.04, 26.19) rectangle (165.26, 46.22);

\path[fill=fillColor] (165.26, 26.19) rectangle (166.49, 45.34);

\path[fill=fillColor] (166.49, 26.19) rectangle (167.71, 45.13);

\path[fill=fillColor] (167.71, 26.19) rectangle (168.94, 44.42);

\path[fill=fillColor] (168.94, 26.19) rectangle (170.17, 44.66);

\path[fill=fillColor] (170.17, 26.19) rectangle (171.39, 45.18);

\path[fill=fillColor] (171.39, 26.19) rectangle (172.62, 44.58);

\path[fill=fillColor] (172.62, 26.19) rectangle (173.85, 44.70);

\path[fill=fillColor] (173.85, 26.19) rectangle (175.07, 45.09);

\path[fill=fillColor] (175.07, 26.19) rectangle (176.30, 44.91);

\path[fill=fillColor] (176.30, 26.19) rectangle (177.52, 44.66);

\path[fill=fillColor] (177.52, 26.19) rectangle (178.75, 45.23);

\path[fill=fillColor] (178.75, 26.19) rectangle (179.98, 44.34);

\path[fill=fillColor] (179.98, 26.19) rectangle (181.20, 44.13);

\path[fill=fillColor] (181.20, 26.19) rectangle (182.43, 44.63);

\path[fill=fillColor] (182.43, 26.19) rectangle (183.66, 44.65);

\path[fill=fillColor] (183.66, 26.19) rectangle (184.88, 43.45);

\path[fill=fillColor] (184.88, 26.19) rectangle (186.11, 44.02);

\path[fill=fillColor] (186.11, 26.19) rectangle (187.34, 43.97);

\path[fill=fillColor] (187.34, 26.19) rectangle (188.56, 44.26);

\path[fill=fillColor] (188.56, 26.19) rectangle (189.79, 43.67);

\path[fill=fillColor] (189.79, 26.19) rectangle (191.01, 44.15);

\path[fill=fillColor] (191.01, 26.19) rectangle (192.24, 43.83);

\path[fill=fillColor] (192.24, 26.19) rectangle (193.47, 43.96);

\path[fill=fillColor] (193.47, 26.19) rectangle (194.69, 44.13);

\path[fill=fillColor] (194.69, 26.19) rectangle (195.92, 44.02);

\path[fill=fillColor] (195.92, 26.19) rectangle (197.15, 44.40);

\path[fill=fillColor] (197.15, 26.19) rectangle (198.37, 43.19);

\path[fill=fillColor] (198.37, 26.19) rectangle (199.60, 43.90);

\path[fill=fillColor] (199.60, 26.19) rectangle (200.82, 44.19);

\path[fill=fillColor] (200.82, 26.19) rectangle (202.05, 43.72);

\path[fill=fillColor] (202.05, 26.19) rectangle (203.28, 43.97);

\path[fill=fillColor] (203.28, 26.19) rectangle (204.50, 44.29);

\path[fill=fillColor] (204.50, 26.19) rectangle (205.73, 43.53);

\path[fill=fillColor] (205.73, 26.19) rectangle (206.96, 43.63);

\path[fill=fillColor] (206.96, 26.19) rectangle (208.18, 43.73);

\path[fill=fillColor] (208.18, 26.19) rectangle (209.41, 44.22);

\path[fill=fillColor] (209.41, 26.19) rectangle (210.63, 43.76);

\path[fill=fillColor] (210.63, 26.19) rectangle (211.86, 43.60);

\path[fill=fillColor] (211.86, 26.19) rectangle (213.09, 44.10);

\path[fill=fillColor] (213.09, 26.19) rectangle (214.31, 45.43);

\path[fill=fillColor] (214.31, 26.19) rectangle (215.54, 45.77);

\path[fill=fillColor] (215.54, 26.19) rectangle (216.77, 45.19);

\path[fill=fillColor] (216.77, 26.19) rectangle (217.99, 45.63);

\path[fill=fillColor] (217.99, 26.19) rectangle (219.22, 50.75);
\definecolor{drawColor}{gray}{0.50}

\path[draw=drawColor,line width= 0.6pt,line join=round,line cap=round] (156.68, 23.76) rectangle (219.22, 77.21);
\end{scope}
\begin{scope}
\path[clip] ( 31.60, 77.21) rectangle ( 94.14, 86.72);
\definecolor{drawColor}{gray}{0.50}
\definecolor{fillColor}{gray}{0.80}

\path[draw=drawColor,line width= 0.2pt,line join=round,line cap=round,fill=fillColor] ( 31.60, 77.21) rectangle ( 94.14, 86.72);
\definecolor{drawColor}{gray}{0.10}

\node[text=drawColor,anchor=base,inner sep=0pt, outer sep=0pt, scale=  0.80] at ( 62.87, 79.21) {fluid height};
\end{scope}
\begin{scope}
\path[clip] ( 94.14, 77.21) rectangle (156.68, 86.72);
\definecolor{drawColor}{gray}{0.50}
\definecolor{fillColor}{gray}{0.80}

\path[draw=drawColor,line width= 0.2pt,line join=round,line cap=round,fill=fillColor] ( 94.14, 77.21) rectangle (156.68, 86.72);
\definecolor{drawColor}{gray}{0.10}

\node[text=drawColor,anchor=base,inner sep=0pt, outer sep=0pt, scale=  0.80] at (125.41, 79.21) {rain content};
\end{scope}
\begin{scope}
\path[clip] (156.68, 77.21) rectangle (219.22, 86.72);
\definecolor{drawColor}{gray}{0.50}
\definecolor{fillColor}{gray}{0.80}

\path[draw=drawColor,line width= 0.2pt,line join=round,line cap=round,fill=fillColor] (156.68, 77.21) rectangle (219.22, 86.72);
\definecolor{drawColor}{gray}{0.10}

\node[text=drawColor,anchor=base,inner sep=0pt, outer sep=0pt, scale=  0.80] at (187.95, 79.21) {wind};
\end{scope}
\begin{scope}
\path[clip] (  0.00,  0.00) rectangle (231.26, 86.72);
\definecolor{drawColor}{RGB}{0,0,0}

\node[text=drawColor,anchor=base east,inner sep=0pt, outer sep=0pt, scale=  0.72] at ( 26.20, 23.71) {0.00};

\node[text=drawColor,anchor=base east,inner sep=0pt, outer sep=0pt, scale=  0.72] at ( 26.20, 33.14) {0.01};

\node[text=drawColor,anchor=base east,inner sep=0pt, outer sep=0pt, scale=  0.72] at ( 26.20, 42.58) {0.02};

\node[text=drawColor,anchor=base east,inner sep=0pt, outer sep=0pt, scale=  0.72] at ( 26.20, 52.01) {0.03};

\node[text=drawColor,anchor=base east,inner sep=0pt, outer sep=0pt, scale=  0.72] at ( 26.20, 61.45) {0.04};

\node[text=drawColor,anchor=base east,inner sep=0pt, outer sep=0pt, scale=  0.72] at ( 26.20, 70.88) {0.05};
\end{scope}
\begin{scope}
\path[clip] (  0.00,  0.00) rectangle (231.26, 86.72);
\definecolor{drawColor}{RGB}{0,0,0}

\path[draw=drawColor,line width= 0.6pt,line join=round] ( 28.60, 26.19) --
	( 31.60, 26.19);

\path[draw=drawColor,line width= 0.6pt,line join=round] ( 28.60, 35.62) --
	( 31.60, 35.62);

\path[draw=drawColor,line width= 0.6pt,line join=round] ( 28.60, 45.06) --
	( 31.60, 45.06);

\path[draw=drawColor,line width= 0.6pt,line join=round] ( 28.60, 54.49) --
	( 31.60, 54.49);

\path[draw=drawColor,line width= 0.6pt,line join=round] ( 28.60, 63.92) --
	( 31.60, 63.92);

\path[draw=drawColor,line width= 0.6pt,line join=round] ( 28.60, 73.36) --
	( 31.60, 73.36);
\end{scope}
\begin{scope}
\path[clip] (  0.00,  0.00) rectangle (231.26, 86.72);
\definecolor{drawColor}{RGB}{0,0,0}

\path[draw=drawColor,line width= 0.6pt,line join=round] ( 43.25, 20.76) --
	( 43.25, 23.76);

\path[draw=drawColor,line width= 0.6pt,line join=round] ( 55.51, 20.76) --
	( 55.51, 23.76);

\path[draw=drawColor,line width= 0.6pt,line join=round] ( 67.77, 20.76) --
	( 67.77, 23.76);

\path[draw=drawColor,line width= 0.6pt,line join=round] ( 80.03, 20.76) --
	( 80.03, 23.76);

\path[draw=drawColor,line width= 0.6pt,line join=round] ( 92.30, 20.76) --
	( 92.30, 23.76);
\end{scope}
\begin{scope}
\path[clip] (  0.00,  0.00) rectangle (231.26, 86.72);
\definecolor{drawColor}{RGB}{0,0,0}

\node[text=drawColor,anchor=base,inner sep=0pt, outer sep=0pt, scale=  0.72] at ( 43.25, 13.40) {10};

\node[text=drawColor,anchor=base,inner sep=0pt, outer sep=0pt, scale=  0.72] at ( 55.51, 13.40) {20};

\node[text=drawColor,anchor=base,inner sep=0pt, outer sep=0pt, scale=  0.72] at ( 67.77, 13.40) {30};

\node[text=drawColor,anchor=base,inner sep=0pt, outer sep=0pt, scale=  0.72] at ( 80.03, 13.40) {40};

\node[text=drawColor,anchor=base,inner sep=0pt, outer sep=0pt, scale=  0.72] at ( 92.30, 13.40) {50};
\end{scope}
\begin{scope}
\path[clip] (  0.00,  0.00) rectangle (231.26, 86.72);
\definecolor{drawColor}{RGB}{0,0,0}

\path[draw=drawColor,line width= 0.6pt,line join=round] (105.79, 20.76) --
	(105.79, 23.76);

\path[draw=drawColor,line width= 0.6pt,line join=round] (118.05, 20.76) --
	(118.05, 23.76);

\path[draw=drawColor,line width= 0.6pt,line join=round] (130.31, 20.76) --
	(130.31, 23.76);

\path[draw=drawColor,line width= 0.6pt,line join=round] (142.58, 20.76) --
	(142.58, 23.76);

\path[draw=drawColor,line width= 0.6pt,line join=round] (154.84, 20.76) --
	(154.84, 23.76);
\end{scope}
\begin{scope}
\path[clip] (  0.00,  0.00) rectangle (231.26, 86.72);
\definecolor{drawColor}{RGB}{0,0,0}

\node[text=drawColor,anchor=base,inner sep=0pt, outer sep=0pt, scale=  0.72] at (105.79, 13.40) {10};

\node[text=drawColor,anchor=base,inner sep=0pt, outer sep=0pt, scale=  0.72] at (118.05, 13.40) {20};

\node[text=drawColor,anchor=base,inner sep=0pt, outer sep=0pt, scale=  0.72] at (130.31, 13.40) {30};

\node[text=drawColor,anchor=base,inner sep=0pt, outer sep=0pt, scale=  0.72] at (142.58, 13.40) {40};

\node[text=drawColor,anchor=base,inner sep=0pt, outer sep=0pt, scale=  0.72] at (154.84, 13.40) {50};
\end{scope}
\begin{scope}
\path[clip] (  0.00,  0.00) rectangle (231.26, 86.72);
\definecolor{drawColor}{RGB}{0,0,0}

\path[draw=drawColor,line width= 0.6pt,line join=round] (168.33, 20.76) --
	(168.33, 23.76);

\path[draw=drawColor,line width= 0.6pt,line join=round] (180.59, 20.76) --
	(180.59, 23.76);

\path[draw=drawColor,line width= 0.6pt,line join=round] (192.85, 20.76) --
	(192.85, 23.76);

\path[draw=drawColor,line width= 0.6pt,line join=round] (205.12, 20.76) --
	(205.12, 23.76);

\path[draw=drawColor,line width= 0.6pt,line join=round] (217.38, 20.76) --
	(217.38, 23.76);
\end{scope}
\begin{scope}
\path[clip] (  0.00,  0.00) rectangle (231.26, 86.72);
\definecolor{drawColor}{RGB}{0,0,0}

\node[text=drawColor,anchor=base,inner sep=0pt, outer sep=0pt, scale=  0.72] at (168.33, 13.40) {10};

\node[text=drawColor,anchor=base,inner sep=0pt, outer sep=0pt, scale=  0.72] at (180.59, 13.40) {20};

\node[text=drawColor,anchor=base,inner sep=0pt, outer sep=0pt, scale=  0.72] at (192.85, 13.40) {30};

\node[text=drawColor,anchor=base,inner sep=0pt, outer sep=0pt, scale=  0.72] at (205.12, 13.40) {40};

\node[text=drawColor,anchor=base,inner sep=0pt, outer sep=0pt, scale=  0.72] at (217.38, 13.40) {50};
\end{scope}
\begin{scope}
\path[clip] (  0.00,  0.00) rectangle (231.26, 86.72);
\definecolor{drawColor}{RGB}{0,0,0}

\node[text=drawColor,anchor=base,inner sep=0pt, outer sep=0pt, scale=  0.90] at (125.41,  2.40) {rank};
\end{scope}
\begin{scope}
\path[clip] (  0.00,  0.00) rectangle (231.26, 86.72);
\definecolor{drawColor}{RGB}{0,0,0}

\node[text=drawColor,rotate= 90.00,anchor=base,inner sep=0pt, outer sep=0pt, scale=  0.90] at (  8.60, 50.49) {frequency};
\end{scope}
\end{tikzpicture}

%% file: LF_boxplot.tex
\begin{tikzpicture}[x=1pt,y=1pt]
\definecolor{fillColor}{RGB}{255,255,255}
\path[use as bounding box,fill=fillColor,fill opacity=0.00] (0,0) rectangle (231.26,108.41);
\begin{scope}
\path[clip] ( 29.60, 42.95) rectangle ( 88.79, 98.90);
\definecolor{fillColor}{RGB}{255,255,255}

\path[fill=fillColor] ( 29.60, 42.95) rectangle ( 88.79, 98.90);
\definecolor{drawColor}{gray}{0.98}

\path[draw=drawColor,line width= 0.6pt,line join=round] ( 29.60, 53.86) --
	( 88.79, 53.86);

\path[draw=drawColor,line width= 0.6pt,line join=round] ( 29.60, 70.59) --
	( 88.79, 70.59);

\path[draw=drawColor,line width= 0.6pt,line join=round] ( 29.60, 87.32) --
	( 88.79, 87.32);
\definecolor{drawColor}{gray}{0.90}

\path[draw=drawColor,line width= 0.2pt,line join=round] ( 29.60, 45.50) --
	( 88.79, 45.50);

\path[draw=drawColor,line width= 0.2pt,line join=round] ( 29.60, 62.22) --
	( 88.79, 62.22);

\path[draw=drawColor,line width= 0.2pt,line join=round] ( 29.60, 78.95) --
	( 88.79, 78.95);

\path[draw=drawColor,line width= 0.2pt,line join=round] ( 29.60, 95.68) --
	( 88.79, 95.68);

\path[draw=drawColor,line width= 0.2pt,line join=round] ( 40.69, 42.95) --
	( 40.69, 98.90);

\path[draw=drawColor,line width= 0.2pt,line join=round] ( 59.19, 42.95) --
	( 59.19, 98.90);

\path[draw=drawColor,line width= 0.2pt,line join=round] ( 77.69, 42.95) --
	( 77.69, 98.90);
\definecolor{drawColor}{RGB}{50,162,81}
\definecolor{fillColor}{RGB}{50,162,81}

\path[draw=drawColor,line width= 0.4pt,line join=round,line cap=round,fill=fillColor] ( 40.69, 72.40) circle (  1.21);

\path[draw=drawColor,line width= 0.4pt,line join=round,line cap=round,fill=fillColor] ( 40.69, 79.11) circle (  1.21);

\path[draw=drawColor,line width= 0.4pt,line join=round,line cap=round,fill=fillColor] ( 40.69, 79.18) circle (  1.21);

\path[draw=drawColor,line width= 0.6pt,line join=round] ( 40.69, 76.56) -- ( 40.69, 78.80);

\path[draw=drawColor,line width= 0.6pt,line join=round] ( 40.69, 74.97) -- ( 40.69, 72.87);
\definecolor{fillColor}{RGB}{50,162,81}

\path[draw=drawColor,line width= 0.6pt,line join=round,line cap=round,fill=fillColor,fill opacity=0.60] ( 33.76, 76.56) --
	( 33.76, 74.97) --
	( 47.63, 74.97) --
	( 47.63, 76.56) --
	( 33.76, 76.56) --
	cycle;

\path[draw=drawColor,line width= 1.1pt,line join=round] ( 33.76, 75.74) -- ( 47.63, 75.74);
\definecolor{drawColor}{RGB}{255,127,15}
\definecolor{fillColor}{RGB}{255,127,15}

\path[draw=drawColor,line width= 0.4pt,line join=round,line cap=round,fill=fillColor] ( 59.19, 82.42) circle (  1.21);

\path[draw=drawColor,line width= 0.4pt,line join=round,line cap=round,fill=fillColor] ( 59.19, 82.82) circle (  1.21);

\path[draw=drawColor,line width= 0.4pt,line join=round,line cap=round,fill=fillColor] ( 59.19, 84.11) circle (  1.21);

\path[draw=drawColor,line width= 0.4pt,line join=round,line cap=round,fill=fillColor] ( 59.19, 92.04) circle (  1.21);

\path[draw=drawColor,line width= 0.4pt,line join=round,line cap=round,fill=fillColor] ( 59.19, 82.63) circle (  1.21);

\path[draw=drawColor,line width= 0.4pt,line join=round,line cap=round,fill=fillColor] ( 59.19, 83.80) circle (  1.21);

\path[draw=drawColor,line width= 0.6pt,line join=round] ( 59.19, 78.21) -- ( 59.19, 81.56);

\path[draw=drawColor,line width= 0.6pt,line join=round] ( 59.19, 75.49) -- ( 59.19, 71.66);
\definecolor{fillColor}{RGB}{255,127,15}

\path[draw=drawColor,line width= 0.6pt,line join=round,line cap=round,fill=fillColor,fill opacity=0.60] ( 52.26, 78.21) --
	( 52.26, 75.49) --
	( 66.13, 75.49) --
	( 66.13, 78.21) --
	( 52.26, 78.21) --
	cycle;

\path[draw=drawColor,line width= 1.1pt,line join=round] ( 52.26, 76.92) -- ( 66.13, 76.92);
\definecolor{drawColor}{RGB}{60,183,204}

\path[draw=drawColor,line width= 0.6pt,line join=round] ( 77.69, 77.52) -- ( 77.69, 80.09);

\path[draw=drawColor,line width= 0.6pt,line join=round] ( 77.69, 75.44) -- ( 77.69, 72.79);
\definecolor{fillColor}{RGB}{60,183,204}

\path[draw=drawColor,line width= 0.6pt,line join=round,line cap=round,fill=fillColor,fill opacity=0.60] ( 70.75, 77.52) --
	( 70.75, 75.44) --
	( 84.63, 75.44) --
	( 84.63, 77.52) --
	( 70.75, 77.52) --
	cycle;

\path[draw=drawColor,line width= 1.1pt,line join=round] ( 70.75, 76.47) -- ( 84.63, 76.47);
\definecolor{drawColor}{gray}{0.50}

\path[draw=drawColor,line width= 0.6pt,line join=round,line cap=round] ( 29.60, 42.95) rectangle ( 88.79, 98.90);
\end{scope}
\begin{scope}
\path[clip] ( 94.81, 42.95) rectangle (154.00, 98.90);
\definecolor{fillColor}{RGB}{255,255,255}

\path[fill=fillColor] ( 94.81, 42.95) rectangle (154.00, 98.90);
\definecolor{drawColor}{gray}{0.98}

\path[draw=drawColor,line width= 0.6pt,line join=round] ( 94.81, 53.86) --
	(154.00, 53.86);

\path[draw=drawColor,line width= 0.6pt,line join=round] ( 94.81, 70.59) --
	(154.00, 70.59);

\path[draw=drawColor,line width= 0.6pt,line join=round] ( 94.81, 87.32) --
	(154.00, 87.32);
\definecolor{drawColor}{gray}{0.90}

\path[draw=drawColor,line width= 0.2pt,line join=round] ( 94.81, 45.50) --
	(154.00, 45.50);

\path[draw=drawColor,line width= 0.2pt,line join=round] ( 94.81, 62.22) --
	(154.00, 62.22);

\path[draw=drawColor,line width= 0.2pt,line join=round] ( 94.81, 78.95) --
	(154.00, 78.95);

\path[draw=drawColor,line width= 0.2pt,line join=round] ( 94.81, 95.68) --
	(154.00, 95.68);

\path[draw=drawColor,line width= 0.2pt,line join=round] (105.91, 42.95) --
	(105.91, 98.90);

\path[draw=drawColor,line width= 0.2pt,line join=round] (124.41, 42.95) --
	(124.41, 98.90);

\path[draw=drawColor,line width= 0.2pt,line join=round] (142.91, 42.95) --
	(142.91, 98.90);
\definecolor{drawColor}{RGB}{50,162,81}
\definecolor{fillColor}{RGB}{50,162,81}

\path[draw=drawColor,line width= 0.4pt,line join=round,line cap=round,fill=fillColor] (105.91, 52.27) circle (  1.21);

\path[draw=drawColor,line width= 0.4pt,line join=round,line cap=round,fill=fillColor] (105.91, 71.21) circle (  1.21);

\path[draw=drawColor,line width= 0.6pt,line join=round] (105.91, 64.19) -- (105.91, 70.18);

\path[draw=drawColor,line width= 0.6pt,line join=round] (105.91, 59.69) -- (105.91, 53.57);
\definecolor{fillColor}{RGB}{50,162,81}

\path[draw=drawColor,line width= 0.6pt,line join=round,line cap=round,fill=fillColor,fill opacity=0.60] ( 98.97, 64.19) --
	( 98.97, 59.69) --
	(112.85, 59.69) --
	(112.85, 64.19) --
	( 98.97, 64.19) --
	cycle;

\path[draw=drawColor,line width= 1.1pt,line join=round] ( 98.97, 61.93) -- (112.85, 61.93);
\definecolor{drawColor}{RGB}{255,127,15}

\path[draw=drawColor,line width= 0.6pt,line join=round] (124.41, 69.57) -- (124.41, 80.14);

\path[draw=drawColor,line width= 0.6pt,line join=round] (124.41, 60.93) -- (124.41, 52.76);
\definecolor{fillColor}{RGB}{255,127,15}

\path[draw=drawColor,line width= 0.6pt,line join=round,line cap=round,fill=fillColor,fill opacity=0.60] (117.47, 69.57) --
	(117.47, 60.93) --
	(131.34, 60.93) --
	(131.34, 69.57) --
	(117.47, 69.57) --
	cycle;

\path[draw=drawColor,line width= 1.1pt,line join=round] (117.47, 65.21) -- (131.34, 65.21);
\definecolor{drawColor}{RGB}{60,183,204}

\path[draw=drawColor,line width= 0.6pt,line join=round] (142.91, 64.55) -- (142.91, 74.54);

\path[draw=drawColor,line width= 0.6pt,line join=round] (142.91, 56.87) -- (142.91, 52.37);
\definecolor{fillColor}{RGB}{60,183,204}

\path[draw=drawColor,line width= 0.6pt,line join=round,line cap=round,fill=fillColor,fill opacity=0.60] (135.97, 64.55) --
	(135.97, 56.87) --
	(149.84, 56.87) --
	(149.84, 64.55) --
	(135.97, 64.55) --
	cycle;

\path[draw=drawColor,line width= 1.1pt,line join=round] (135.97, 60.06) -- (149.84, 60.06);
\definecolor{drawColor}{gray}{0.50}

\path[draw=drawColor,line width= 0.6pt,line join=round,line cap=round] ( 94.81, 42.95) rectangle (154.00, 98.90);
\end{scope}
\begin{scope}
\path[clip] (160.03, 42.95) rectangle (219.22, 98.90);
\definecolor{fillColor}{RGB}{255,255,255}

\path[fill=fillColor] (160.03, 42.95) rectangle (219.22, 98.90);
\definecolor{drawColor}{gray}{0.98}

\path[draw=drawColor,line width= 0.6pt,line join=round] (160.03, 53.86) --
	(219.22, 53.86);

\path[draw=drawColor,line width= 0.6pt,line join=round] (160.03, 70.59) --
	(219.22, 70.59);

\path[draw=drawColor,line width= 0.6pt,line join=round] (160.03, 87.32) --
	(219.22, 87.32);
\definecolor{drawColor}{gray}{0.90}

\path[draw=drawColor,line width= 0.2pt,line join=round] (160.03, 45.50) --
	(219.22, 45.50);

\path[draw=drawColor,line width= 0.2pt,line join=round] (160.03, 62.22) --
	(219.22, 62.22);

\path[draw=drawColor,line width= 0.2pt,line join=round] (160.03, 78.95) --
	(219.22, 78.95);

\path[draw=drawColor,line width= 0.2pt,line join=round] (160.03, 95.68) --
	(219.22, 95.68);

\path[draw=drawColor,line width= 0.2pt,line join=round] (171.12, 42.95) --
	(171.12, 98.90);

\path[draw=drawColor,line width= 0.2pt,line join=round] (189.62, 42.95) --
	(189.62, 98.90);

\path[draw=drawColor,line width= 0.2pt,line join=round] (208.12, 42.95) --
	(208.12, 98.90);
\definecolor{drawColor}{RGB}{50,162,81}
\definecolor{fillColor}{RGB}{50,162,81}

\path[draw=drawColor,line width= 0.4pt,line join=round,line cap=round,fill=fillColor] (171.12, 72.03) circle (  1.21);

\path[draw=drawColor,line width= 0.4pt,line join=round,line cap=round,fill=fillColor] (171.12, 70.31) circle (  1.21);

\path[draw=drawColor,line width= 0.4pt,line join=round,line cap=round,fill=fillColor] (171.12, 72.47) circle (  1.21);

\path[draw=drawColor,line width= 0.6pt,line join=round] (171.12, 77.57) -- (171.12, 79.21);

\path[draw=drawColor,line width= 0.6pt,line join=round] (171.12, 75.56) -- (171.12, 72.97);
\definecolor{fillColor}{RGB}{50,162,81}

\path[draw=drawColor,line width= 0.6pt,line join=round,line cap=round,fill=fillColor,fill opacity=0.60] (164.19, 77.57) --
	(164.19, 75.56) --
	(178.06, 75.56) --
	(178.06, 77.57) --
	(164.19, 77.57) --
	cycle;

\path[draw=drawColor,line width= 1.1pt,line join=round] (164.19, 76.83) -- (178.06, 76.83);
\definecolor{drawColor}{RGB}{255,127,15}
\definecolor{fillColor}{RGB}{255,127,15}

\path[draw=drawColor,line width= 0.4pt,line join=round,line cap=round,fill=fillColor] (189.62, 72.48) circle (  1.21);

\path[draw=drawColor,line width= 0.4pt,line join=round,line cap=round,fill=fillColor] (189.62, 73.05) circle (  1.21);

\path[draw=drawColor,line width= 0.4pt,line join=round,line cap=round,fill=fillColor] (189.62, 81.58) circle (  1.21);

\path[draw=drawColor,line width= 0.4pt,line join=round,line cap=round,fill=fillColor] (189.62, 84.81) circle (  1.21);

\path[draw=drawColor,line width= 0.4pt,line join=round,line cap=round,fill=fillColor] (189.62, 69.46) circle (  1.21);

\path[draw=drawColor,line width= 0.6pt,line join=round] (189.62, 78.57) -- (189.62, 81.28);

\path[draw=drawColor,line width= 0.6pt,line join=round] (189.62, 76.59) -- (189.62, 73.78);
\definecolor{fillColor}{RGB}{255,127,15}

\path[draw=drawColor,line width= 0.6pt,line join=round,line cap=round,fill=fillColor,fill opacity=0.60] (182.69, 78.57) --
	(182.69, 76.59) --
	(196.56, 76.59) --
	(196.56, 78.57) --
	(182.69, 78.57) --
	cycle;

\path[draw=drawColor,line width= 1.1pt,line join=round] (182.69, 77.52) -- (196.56, 77.52);
\definecolor{drawColor}{RGB}{60,183,204}
\definecolor{fillColor}{RGB}{60,183,204}

\path[draw=drawColor,line width= 0.4pt,line join=round,line cap=round,fill=fillColor] (208.12, 71.95) circle (  1.21);

\path[draw=drawColor,line width= 0.4pt,line join=round,line cap=round,fill=fillColor] (208.12, 72.90) circle (  1.21);

\path[draw=drawColor,line width= 0.4pt,line join=round,line cap=round,fill=fillColor] (208.12, 69.65) circle (  1.21);

\path[draw=drawColor,line width= 0.4pt,line join=round,line cap=round,fill=fillColor] (208.12, 72.51) circle (  1.21);

\path[draw=drawColor,line width= 0.6pt,line join=round] (208.12, 77.86) -- (208.12, 79.45);

\path[draw=drawColor,line width= 0.6pt,line join=round] (208.12, 75.89) -- (208.12, 72.95);
\definecolor{fillColor}{RGB}{60,183,204}

\path[draw=drawColor,line width= 0.6pt,line join=round,line cap=round,fill=fillColor,fill opacity=0.60] (201.18, 77.86) --
	(201.18, 75.89) --
	(215.06, 75.89) --
	(215.06, 77.86) --
	(201.18, 77.86) --
	cycle;

\path[draw=drawColor,line width= 1.1pt,line join=round] (201.18, 77.06) -- (215.06, 77.06);
\definecolor{drawColor}{gray}{0.50}

\path[draw=drawColor,line width= 0.6pt,line join=round,line cap=round] (160.03, 42.95) rectangle (219.22, 98.90);
\end{scope}
\begin{scope}
\path[clip] ( 29.60, 98.90) rectangle ( 88.79,108.41);
\definecolor{drawColor}{gray}{0.50}
\definecolor{fillColor}{gray}{0.80}

\path[draw=drawColor,line width= 0.2pt,line join=round,line cap=round,fill=fillColor] ( 29.60, 98.90) rectangle ( 88.79,108.41);
\definecolor{drawColor}{gray}{0.10}

\node[text=drawColor,anchor=base,inner sep=0pt, outer sep=0pt, scale=  0.80] at ( 59.19,100.90) {fluid height};
\end{scope}
\begin{scope}
\path[clip] ( 94.81, 98.90) rectangle (154.00,108.41);
\definecolor{drawColor}{gray}{0.50}
\definecolor{fillColor}{gray}{0.80}

\path[draw=drawColor,line width= 0.2pt,line join=round,line cap=round,fill=fillColor] ( 94.81, 98.90) rectangle (154.00,108.41);
\definecolor{drawColor}{gray}{0.10}

\node[text=drawColor,anchor=base,inner sep=0pt, outer sep=0pt, scale=  0.80] at (124.41,100.90) {rain content};
\end{scope}
\begin{scope}
\path[clip] (160.03, 98.90) rectangle (219.22,108.41);
\definecolor{drawColor}{gray}{0.50}
\definecolor{fillColor}{gray}{0.80}

\path[draw=drawColor,line width= 0.2pt,line join=round,line cap=round,fill=fillColor] (160.03, 98.90) rectangle (219.22,108.41);
\definecolor{drawColor}{gray}{0.10}

\node[text=drawColor,anchor=base,inner sep=0pt, outer sep=0pt, scale=  0.80] at (189.62,100.90) {wind};
\end{scope}
\begin{scope}
\path[clip] (  0.00,  0.00) rectangle (231.26,108.41);
\definecolor{drawColor}{RGB}{0,0,0}

\node[text=drawColor,anchor=base east,inner sep=0pt, outer sep=0pt, scale=  0.72] at ( 24.20, 43.02) {50};

\node[text=drawColor,anchor=base east,inner sep=0pt, outer sep=0pt, scale=  0.72] at ( 24.20, 59.75) {75};

\node[text=drawColor,anchor=base east,inner sep=0pt, outer sep=0pt, scale=  0.72] at ( 24.20, 76.47) {100};

\node[text=drawColor,anchor=base east,inner sep=0pt, outer sep=0pt, scale=  0.72] at ( 24.20, 93.20) {125};
\end{scope}
\begin{scope}
\path[clip] (  0.00,  0.00) rectangle (231.26,108.41);
\definecolor{drawColor}{RGB}{0,0,0}

\path[draw=drawColor,line width= 0.6pt,line join=round] ( 26.60, 45.50) --
	( 29.60, 45.50);

\path[draw=drawColor,line width= 0.6pt,line join=round] ( 26.60, 62.22) --
	( 29.60, 62.22);

\path[draw=drawColor,line width= 0.6pt,line join=round] ( 26.60, 78.95) --
	( 29.60, 78.95);

\path[draw=drawColor,line width= 0.6pt,line join=round] ( 26.60, 95.68) --
	( 29.60, 95.68);
\end{scope}
\begin{scope}
\path[clip] (  0.00,  0.00) rectangle (231.26,108.41);
\definecolor{drawColor}{RGB}{0,0,0}

\path[draw=drawColor,line width= 0.6pt,line join=round] ( 40.69, 39.95) --
	( 40.69, 42.95);

\path[draw=drawColor,line width= 0.6pt,line join=round] ( 59.19, 39.95) --
	( 59.19, 42.95);

\path[draw=drawColor,line width= 0.6pt,line join=round] ( 77.69, 39.95) --
	( 77.69, 42.95);
\end{scope}
\begin{scope}
\path[clip] (  0.00,  0.00) rectangle (231.26,108.41);
\definecolor{drawColor}{RGB}{0,0,0}

\node[text=drawColor,rotate= 45.00,anchor=base east,inner sep=0pt, outer sep=0pt, scale=  0.72] at ( 44.20, 34.05) {\textsc{le}n\textsc{kf}};

\node[text=drawColor,rotate= 45.00,anchor=base east,inner sep=0pt, outer sep=0pt, scale=  0.72] at ( 62.70, 34.05) {\textsc{naive-le}n\textsc{kpf}};

\node[text=drawColor,rotate= 45.00,anchor=base east,inner sep=0pt, outer sep=0pt, scale=  0.72] at ( 81.20, 34.05) {\textsc{block-le}n\textsc{kpf}};
\end{scope}
\begin{scope}
\path[clip] (  0.00,  0.00) rectangle (231.26,108.41);
\definecolor{drawColor}{RGB}{0,0,0}

\path[draw=drawColor,line width= 0.6pt,line join=round] (105.91, 39.95) --
	(105.91, 42.95);

\path[draw=drawColor,line width= 0.6pt,line join=round] (124.41, 39.95) --
	(124.41, 42.95);

\path[draw=drawColor,line width= 0.6pt,line join=round] (142.91, 39.95) --
	(142.91, 42.95);
\end{scope}
\begin{scope}
\path[clip] (  0.00,  0.00) rectangle (231.26,108.41);
\definecolor{drawColor}{RGB}{0,0,0}

\node[text=drawColor,rotate= 45.00,anchor=base east,inner sep=0pt, outer sep=0pt, scale=  0.72] at (109.42, 34.05) {\textsc{le}n\textsc{kf}};

\node[text=drawColor,rotate= 45.00,anchor=base east,inner sep=0pt, outer sep=0pt, scale=  0.72] at (127.91, 34.05) {\textsc{naive-le}n\textsc{kpf}};

\node[text=drawColor,rotate= 45.00,anchor=base east,inner sep=0pt, outer sep=0pt, scale=  0.72] at (146.41, 34.05) {\textsc{block-le}n\textsc{kpf}};
\end{scope}
\begin{scope}
\path[clip] (  0.00,  0.00) rectangle (231.26,108.41);
\definecolor{drawColor}{RGB}{0,0,0}

\path[draw=drawColor,line width= 0.6pt,line join=round] (171.12, 39.95) --
	(171.12, 42.95);

\path[draw=drawColor,line width= 0.6pt,line join=round] (189.62, 39.95) --
	(189.62, 42.95);

\path[draw=drawColor,line width= 0.6pt,line join=round] (208.12, 39.95) --
	(208.12, 42.95);
\end{scope}
\begin{scope}
\path[clip] (  0.00,  0.00) rectangle (231.26,108.41);
\definecolor{drawColor}{RGB}{0,0,0}

\node[text=drawColor,rotate= 45.00,anchor=base east,inner sep=0pt, outer sep=0pt, scale=  0.72] at (174.63, 34.05) {\textsc{le}n\textsc{kf}};

\node[text=drawColor,rotate= 45.00,anchor=base east,inner sep=0pt, outer sep=0pt, scale=  0.72] at (193.13, 34.05) {\textsc{naive-le}n\textsc{kpf}};

\node[text=drawColor,rotate= 45.00,anchor=base east,inner sep=0pt, outer sep=0pt, scale=  0.72] at (211.63, 34.05) {\textsc{block-le}n\textsc{kpf}};
\end{scope}
\begin{scope}
\path[clip] (  0.00,  0.00) rectangle (231.26,108.41);
\definecolor{drawColor}{RGB}{0,0,0}

\node[text=drawColor,rotate= 90.00,anchor=base,inner sep=0pt, outer sep=0pt, scale=  0.90] at (  8.60, 70.92) {\textsc{crps}};
\end{scope}
\end{tikzpicture}